\documentclass[sigconf]{acmart}

\renewcommand\footnotetextcopyrightpermission[1]{} 
\usepackage{booktabs} 
\usepackage{multirow}
\usepackage{booktabs}
\usepackage{graphicx}  
\usepackage{diagbox}
\usepackage[braket]{qcircuit} 
\usepackage{threeparttable} 
\usepackage{stfloats} 
\usepackage{enumitem} 
\newcommand{\tabincell}[2]{\begin{tabular}{@{}#1@{}}#2\end{tabular}}

\setcopyright{none}

\acmArticle{4}

\begin{document}
\title{Quantum Software Engineering\\ {\LARGE\it Landscapes and Horizons}}

\author{Jianjun Zhao}

\affiliation{%
  \institution{Kyushu University}
}
\email{zhao@ait.kyushu-u.ac.jp}

\renewcommand{\shortauthors}{J. Zhao}

\begin{abstract}
Quantum software plays a critical role in exploiting the full potential of quantum computing systems. As a result, it has been drawing increasing attention recently. This paper defines the term “quantum software engineering” and introduces a quantum software life cycle. The paper also gives a generic view of quantum software engineering and discusses the quantum software engineering processes, methods, and tools. Based on these, the paper provides a comprehensive survey of the current state of the art in the field and presents the challenges and opportunities we face. The survey summarizes the technology available in the various phases of the quantum software life cycle, including quantum software requirements analysis, design, implementation, test, and maintenance. It also covers the crucial issues of quantum software reuse and measurement.
\end{abstract}


\maketitle

\section{Introduction}
\label{sec:introduction}

\vspace*{3mm}
{\it “The challenge [of quantum software engineering] is to rework and extend the whole of classical software engineering into the quantum domain so that programmers can manipulate quantum programs with the same ease and confidence that they manipulate today's classical programs.”}
\begin{flushright} 
\hspace*{1cm}{\bf Susan Stepney}, in the 2004 report of Grand Challenges in Computing Research~\cite{hoare2004grand}.
\end{flushright}
\vspace*{2mm}

Quantum computing is a rapidly developing field that is expected to make breakthroughs in many areas~\cite{lanyon2010towards,barends2014superconducting,cross2015quantum,benedetti2016estimation,o2016scalable,olson2017quantum}. Quantum computing uses the principles of quantum mechanics to process information and has the potential to perform specific tasks much faster than classical computing. The basis of quantum computing is the quantum bit (qubit). Unlike a classical computer bit that is assigned either 0 or 1, a qubit can be either assigned states that is the superposition of 0 and 1. Compared to classical algorithms, quantum algorithms have the potential to solve specific problems with exponential acceleration. In the last two decades, there has been a large amount of literature that contributed to the development of quantum algorithms~\cite{deutsch1985quantum,grover1996fast,shor1999polynomial,mosca2008quantum,montanaro2016quantum,shao2019quantum}.

Just like classical computing, quantum computing is applicable to applications in many disciplines and will have a wide impact~\cite{glanz1995quantum}. There are two types of applications where quantum computers are expected to outperform classical computers~\cite{martonosi2019next}. The first application is the problems that require large amounts of parallel computing, such as optimization~\cite{guerreschi2017practical,farhi2014quantum}, encryption~\cite{mosca2018cybersecurity}, big data analysis~\cite{rebentrost2014quantum}, and machine learning~\cite{dunjko2016quantum,biamonte2017quantum}. The other application is the problems that require efficient and accurate simulation of quantum problems in nature~\cite{feynman1982simulating,zalka1998efficient} from the areas, such as physics~\cite{childs2018toward}, chemistry~\cite{reiher2017elucidating,mcardle2018quantum,olson2017quantum} and materials science~\cite{yang2017mixed,grimsley2019adaptive}.

With the rapid development of impressive quantum hardware as well as the accessibility of universal quantum devices to the researchers and professionals via Quantum-as-a-Service (QaaS)~\cite{leymann2020quantum,ball2021software}, it is high time to focus our attention on engineering quantum software systems to reap the benefits of quantum technology. Meanwhile, various application domains of quantum computing~\cite{wecker2013can,wecker2014gate,karalekas2020quantum,tura2020quantum} urgently need quantum software engineering methodologies and techniques to support solving their specific problems. Therefore, we believe that there is an urgent need to build a community for quantum software engineering that focuses on devising methods, tools, and processes for developing quantum software systems efficiently.

Quantum software development techniques aim at providing means for creating quantum software throughout the quantum software life cycle (refer to Section~\ref{subsec:QSLC}). A number of quantum programming approaches are available, for instance, Scaffold~\cite{abhari2012scaffold,javadiabhari2015scaffcc}, Qiskit~\cite{ibm2017qiskit}, Q\#~\cite{svore2018q}, ProjectQ~\cite{projectq2017projectq}, and Quipper\cite{green2013quipper}. Moreover, the concepts are also being applied at the earlier stages of quantum software development. For example, at the quantum software design stage, \cite{Perez-Delgado2020quantum,carmelo2013quantum,ali2020modeling} provides means for modeling and specifying quantum software systems, and \cite{thompson2018quantum,sanchez2021definition} studied the modularity issue in the design of quantum systems. With a variety of techniques available to a software engineer at each stage, the task of engineering a quantum software system poses significant challenges. At each development stage, the software engineer needs to employ the most suitable quantum software technique for the application being developed. The choice of technique can be dictated by various factors, including system requirements, organizational practices, and constraints imposed by the tools or development environments, and the nature of the quantum programming (mechanics). This implies that multiple techniques may be employed at each stage in conjunction with each other. As quantum software development techniques mature, there is a need for guidelines supporting the development of well-engineered quantum software systems. 

In this paper, we present a comprehensive survey of quantum software engineering. We summarize the aspects of previous work that have focused mainly on quantum software engineering issues, while simultaneously covering various approaches across different phases of the quantum software life cycle. We have organized the literature according to five different aspects: requirements, design, implementation, testing, and maintenance. Some papers may cover more than one aspect. For such papers, we refer them to all the relevant sections to ensure completeness.

Additionally, we identify challenges and opportunities for the emerging research community working at the intersection between methodologies and techniques for classical software engineering and problems in engineering quantum software systems. To ensure that our survey is self-contained, we have tried to include enough materials to thoroughly guide the software engineering researchers who are interested in techniques and methods for engineering quantum software systems.
We also seek to provide quantum computing researchers with a complete survey of software engineering solutions to improve the development of quantum software systems.

There have been surveys related to some aspects of quantum software engineering. For example, a number of surveys~\cite{selinger2004brief,gay2006quantum,unruh2006quantum,rudiger2007quantum,jorrand2007programmer, sofge2008survey,miszczak2011models,ying2012quantum,valiron2013quantum,valiron2015programming,hietala2016quantum,chong2017programming,spivsiak2017quantum,zorzi2019quantum,garhwal2019quantum} have been proposed for quantum programming languages, which mainly focus on the language paradigms, implementations, semantics, and tools. Surveys related to quantum software testing~\cite{garcia2021quantum}, quantum program verification~\cite{lewis2021formal,chareton2021formal}, and quantum software development environments~\cite{roetteler2017design,larose2019overview, fingerhuth2018open, shaydulin2020making} have also been proposed, from various points of view. However, as far as we know, no previous work has focused on a comprehensive survey of the whole life cycle of quantum software development, including quantum software requirements analysis, design, implementation, testing, and maintenance. 

In summary, the main contributions of this paper are as follows:

\begin{itemize}
\item{\bf Definition.} It defines the term “quantum software engineering” and introduces a quantum software life cycle model for supporting quantum software development.
\item{\bf Survey.} It provides a comprehensive survey on quantum software engineering, which is across various phases of quantum software life cycle, including quantum software requirements analysis, design, implementation, testing, and maintenance.
\item{\bf Horizons.} It identifies challenges, opportunities, and promising research directions for quantum software engineering, intended to promote and stimulate further research.
\end{itemize}

The rest of this paper is organized as follows. The structure of this paper is depicted in Table~\ref{table:structure}.  
Section~\ref{sec:background} reviews the fundamental terminology in quantum computing. Section~\ref{sec:QSD} briefly introduces the quantum software development which includes quantum programming, a definition of “quantum software engineering,” and the quantum software life cycle for quantum software development. From Section~\ref{sec:requirement} to Section~\ref{sec:maintenance}, we survey the current state of the art on quantum software requirements analysis, design, implementation, testing, and maintenance, respectively. Section~\ref{sec:reuse} and Section~\ref{sec:measurement} cover the current state of the art on quantum software reuse and measurement, respectively. Section~\ref{sec:challenge} discusses the challenges and opportunities for quantum software engineering. Related work is discussed in Section \ref{sec:work}, and concluding remarks are given in Section~\ref{sec:conclusion}. An extensive set of references is provided for readers wishing to pursue the matter further.

\begin{table*}[h]
\centering
\caption{A Table-like Structure for Paper Organization}
\label{table:structure}
\renewcommand\arraystretch{1.2}
\scriptsize
\begin{tabular}{|l|l|l|} 
\hline 
{\bf Introduction} \hspace{0.3mm} [Sec.\ref{sec:introduction}] & \multicolumn{2}{l|}{}  \\
\hline 
\multirow{9}{*}{{\bf Background} \hspace{0.3mm} [Sec.\ref{sec:background}]} 
                        & \multicolumn{2}{l|}{Quantum bit (qubit) \hspace{0.3mm} [Sec.\ref{subsec:qubit}]}  \\\cline{2-3}
                        & \multirow{3}{*}{Quantum gate \hspace{0.3mm} [Sec.\ref{subsec:q-gate}]} 
                                                   & NOT gate \hspace{0.3mm} [Sec.\ref{subsubsec:not}] \\\cline{3-3}
                        &                          & Hadamard gate \hspace{0.3mm} [Sec.\ref{subsubsec:hadamard}] \\\cline{3-3}
                        &                          & Controlled NOT gate \hspace{0.3mm} [Sec.\ref{subsubsec:controlled}] \\\cline{2-3}

                        & \multicolumn{2}{l|}{Quantum circuit \hspace{0.3mm} [Sec.\ref{subsec:q-circuit}]}  \\\cline{2-3}
                        & \multicolumn{2}{l|}{Superposition and entanglement \hspace{0.3mm} [Sec.\ref{subsec:q-entanglement}]}  \\\cline{2-3}       
                        & \multicolumn{2}{l|}{Quantum measurement \hspace{0.3mm} [Sec.\ref{subsec:q-measurement}]}  \\\cline{2-3}
                        & \multicolumn{2}{l|}{Quantum algorithm \hspace{0.3mm} [Sec.\ref{subsec:q-algorithm}]}  \\\cline{2-3}    
                        & \multicolumn{2}{l|}{Quantum computing \hspace{0.3mm} [Sec.\ref{subsec:q-computing}]}  \\    
\hline
\multirow{7}{*}{\tabincell{l}{{\bf Quantum Software Engineering} \hspace{0.3mm} [Sec.\ref{sec:QSD}]}} & \multirow{3}{*}{\tabincell{l}{Quantum programming \hspace{0.3mm} [Sec.\ref{subsec:q-programming}]}} & Concepts of quantum programming \hspace{0.3mm} [Sec.\ref{subsubsec:concept}] \\\cline{3-3}
                         &                          & Languages for quantum programming \hspace{0.3mm} [Sec.\ref{subsubsec:QPL-qpl}] \\\cline{3-3}
                         &                          & Semantics of quantum programming \hspace{0.3mm} [Sec.\ref{subsubsec:QPL-semantics}] \\\cline{2-3}

                                        & \multicolumn{2}{l|}{Definition of quantum software engineering \hspace{0.3mm} [Sec.\ref{subsec:QSE-definition}]}  \\\cline{2-3}
                                        & \multicolumn{2}{l|}{Quantum software engineering methods, tools, and processes \hspace{0.3mm} [Sec.\ref{subsec:QSE-process}]}  \\\cline{2-3}
                                        & \multicolumn{2}{l|}{Generic view of quantum software engineering \hspace{0.3mm} [Sec.\ref{subsec:QSE-view}]}  \\\cline{2-3}
                                        & \multicolumn{2}{l|}{Quantum software life cycle \hspace{0.3mm} [Sec.\ref{subsec:QSLC}]}  \\            
\hline
{\bf Quantum Requirements Analysis} \hspace{0.3mm} [Sec.\ref{sec:requirement}] & \multicolumn{2}{l|}{}  \\
\hline
\multirow{4}{*}{{\bf Quantum Software Design} \hspace{0.3mm} [Sec.\ref{sec:design}]} & \multicolumn{2}{l|}{Quantum software modeling \hspace{0.3mm} [Sec.\ref{subsec:d-modelling}]} \\\cline{2-3}
                                        & \multicolumn{2}{l|}{Quantum software specification \hspace{0.3mm} [Sec.\ref{subsec:d-specification}]}  \\\cline{2-3}              
                                        & \multicolumn{2}{l|}{Pattern language for quantum software \hspace{0.3mm} [Sec.\ref{subsec:q-pattern}]}  \\\cline{2-3}              & \multicolumn{2}{l|}{Modular design for quantum systems \hspace{0.3mm} [Sec.\ref{subsec:modular-design}]}  \\ 
\hline
{\bf Quantum Software Implementation} \hspace{0.3mm} [Sec.\ref{sec:implementation}] &  \multicolumn{2}{l|}{}   \\
\hline
\multirow{14}{*}{{\bf Quantum Software Testing} \hspace{0.3mm} [Sec.\ref{sec:testing}]} 
                         & \multirow{2}{*}{Bugs in quantum software \hspace{0.3mm} [Sec.\ref{subsec:bugs}]} & Bug types and patterns \hspace{0.3mm} [Sec.\ref{subsubsec:bug-pattern}] \\\cline{3-3}
                         &                          & Bug benchmarks \hspace{0.3mm} [Sec.\ref{subsubsec:bug-benchmark}] \\\cline{2-3}
                         & \multirow{3}{*}{Assertions for quantum software \hspace{0.3mm} [Sec.\ref{subsec:q-assertion}]} & Invariant and inductive assertion \hspace{0.3mm} [Sec.\ref{subsubsec:invariant}] \\\cline{3-3}
                         &                          & Applied quantum Hoare logic \hspace{0.3mm} [Sec.\ref{subsubsec:aQHL}] \\\cline{3-3}
                         &                          & Assertion library for quantum software \hspace{0.3mm} [Sec.\ref{subsubsec:property}] \\\cline{2-3}
                         & \multirow{4}{*}{Quantum software testing \hspace{0.3mm} [Sec.\ref{subsec:q-testing}]} & Challenges on quantum software testing \hspace{0.3mm} [Sec.\ref{subsubsec:open-problem}] \\\cline{3-3}
                         &                          & Testing approaches \hspace{0.3mm} [Sec.\ref{subsubsec:testing-approach}] \\\cline{3-3}
                         &                          & Testing adequacy criteria \hspace{0.3mm} [Sec.\ref{subsubsec:testing-criteria}] \\\cline{2-3}
                         & \multirow{4}{*}{Quantum program debugging \hspace{0.3mm} [Sec.\ref{subsec:q-debugging}]} & Debugging tactics \hspace{0.3mm} [Sec.\ref{subsubsec:debugging-tactic}] \\\cline{3-3}
                         &                            & Assertion-based debugging \hspace{0.3mm} [Sec.\ref{subsubsec:assertion-debugging}]\\\cline{3-3}
                         &                            & Debugging quantum processes \hspace{0.3mm} [Sec.\ref{subsubsec:quantum-process}] \\\cline{3-3}
                         &                            & Language support for debugging \hspace{0.3mm} [Sec.\ref{subsubsec:debugging-language}] \\\cline{2-3}
                         & \multirow{2}{*}{Quantum program analysis \hspace{0.3mm} [Sec.\ref{subsec:q-analysis}]} & Entanglement analysis \hspace{0.3mm} [Sec.\ref{subsubsec:entanglement-analysis}] \\\cline{3-3}
                         &                          & Robustness analysis \hspace{0.3mm} [Sec.\ref{subsubsec:robustness-analysis}] \\\cline{2-3}
\hline
\tabincell{l}{{\bf Quantum Software Maintenance} \hspace{0.3mm} [Sec.\ref{sec:maintenance}]} &  \multicolumn{2}{l|}{Reengineering classical information system to quantum computing \hspace{0.3mm} [Sec.\ref{subsec:reengineering}]} \\
\hline
\multirow{3}{*}{\tabincell{l}{{\bf Quantum Software Reuse} \hspace{0.3mm} [Sec.\ref{sec:reuse}]}} & \multicolumn{2}{l|}{Quantum pattern reuse \hspace{0.3mm} [Sec.\ref{subsec:pattern-reuse}]} \\\cline{2-3}
                                        & \multicolumn{2}{l|}{Quantum circuit reuse \hspace{0.3mm} [Sec.\ref{subsec:circuit-reuse}]} \\\cline{2-3}
                                        & \multicolumn{2}{l|}{Quantum state reuse \hspace{0.3mm} [Sec.\ref{subsec:state-reuse}]} \\
\hline
\multirow{2}{*}{\tabincell{l}{{\bf Quantum Software Measurement} \hspace{0.3mm} [Sec.\ref{sec:measurement}]}} & \multicolumn{2}{l|}{Size and structure metrics \hspace{0.3mm} [Sec.\ref{subsec:size-metric}]} \\\cline{2-3}
                                        & \multicolumn{2}{l|}{Metrics for quantum circuits understandability \hspace{0.3mm} [Sec.\ref{subsubsec:circuit-metric}]} \\
\hline
\tabincell{l}{{\bf Empirical Study on Quantum Software Engineering} \hspace{0.3mm} [Sec.\ref{sec:empirical-study}]} &  \multicolumn{2}{l|}{} \\
\hline
\multirow{3}{*}{\tabincell{l}{{\bf Software Engineering for Quantum Computing Platforms} \hspace{0.3mm} [Sec.\ref{sec:quantum-qcp}]}} & \multicolumn{2}{l|}{Quality attributes on quantum computing platforms \hspace{0.3mm} [Sec.\ref{subsec:attribute-qcp}]} \\\cline{2-3}
                                        & \multicolumn{2}{l|}{Bugs in quantum computing platforms \hspace{0.3mm} [Sec.\ref{subsec:bug-qcp}]} \\\cline{2-3}
                                        & \multicolumn{2}{l|}{Testing quantum software stacks \hspace{0.3mm} [Sec.\ref{subsec:testing-qcp}]} \\
\hline
\multirow{12}{*}{\tabincell{l}{{\bf Challenge and Opportunities} \hspace{0.3mm} [Sec.\ref{sec:challenge}]}} & \multicolumn{2}{l|}{Quantum software requirements analysis \hspace{0.3mm} [Sec.\ref{subsec:co-requirement}]} \\\cline{2-3}
                         &\multirow{2}{*}{\tabincell{l}{{Quantum software design} \hspace{0.3mm} [Sec.\ref{subsec:co-design}]}} & Quantum architectural design \hspace{0.3mm} [Sec.\ref{subsubsec:co-architectural}] \\\cline{3-3}
                         &                          & Detailed quantum design \hspace{0.3mm} [Sec.\ref{subsubsec:co-detail}] \\\cline{3-3}
                         &                          & Design models for quantum software \hspace{0.3mm} [Sec.\ref{subsubsec:co-model}] \\\cline{2-3}      
                         & \multicolumn{2}{l|}{Quantum software implementation \hspace{0.3mm} [Sec.\ref{subsec:co-implementation}]}  \\\cline{2-3}          
                         & \multirow{5}{*}{Quantum software reliability [Sec.\ref{subsec:co-testing}]} & Fault model for quantum software \hspace{0.3mm} [Sec.\ref{subsubsec:co-fault}] \\\cline{3-3}
                         &                          & Quantum software testing \hspace{0.3mm} [Sec.\ref{subsubsec:co-testing}] \\\cline{3-3}
                         &                          & Quantum program debugging \hspace{0.3mm} [Sec.\ref{subsubsec:co-debugging}] \\\cline{3-3}
                         &                          & Quantum software visualization \hspace{0.3mm} [Sec.\ref{subsubsec:co-visualization}] \\\cline{3-3}
                         &                          & Quantum program verification \hspace{0.3mm} [Sec.\ref{subsubsec:co-verification}] \\\cline{2-3}
                         & \multicolumn{2}{l|}{Quantum software maintenance \hspace{0.3mm} [Sec.\ref{subsec:co-maintenance}]}  \\\cline{2-3}          
                         & \multicolumn{2}{l|}{Quantum software resue \hspace{0.3mm} [Sec.\ref{subsec:co-reuse}]}  \\            
\hline 
\multirow{4}{*}{{\bf Related work} \hspace{0.3mm} [Sec.\ref{sec:work}]} & \multicolumn{2}{l|}{Quantum programming languages \hspace{0.3mm} [Sec.\ref{subsec:QPL-survey}]} \\\cline{2-3}
                                        & \multicolumn{2}{l|}{Quantum software engineering \hspace{0.3mm}[Sec.\ref{subsec:QSE}]}  \\\cline{2-3}
                                        & \multicolumn{2}{l|}{Quantum software reliability \hspace{0.3mm} [Sec.\ref{subsec:q-reliability}]}  \\\cline{2-3}         & \multicolumn{2}{l|}{Quantum software development environments \hspace{0.3mm} [Sec.\ref{subsec:IDE}]}  \\
\hline
{\bf Concluding Remarks} \hspace{0.3mm} [Sec.\ref{sec:conclusion}] & \multicolumn{2}{l|}{}  \\
\hline
\end{tabular}
\end{table*}

\section{Background}
\label{sec:background}

This section briefly introduces some basics of quantum mechanics, which form the basis of quantum computing. More detailed reading materials can be found in the books by Gruska~\cite{gruska1999quantum}, Nielsen \& Chuang~\cite{nielsen2002quantum}, and Nakahara~\cite{nakahara2008quantum}. Preskill's lecture notes~\cite{preskill2018lecture} are also a valuable resource. 

\subsection{Quantum Bit (Qubit)}
\label{subsec:qubit}

A classical bit is a binary unit of information used in classical computation. It can take two possible values, 0 or 1. A quantum bit (or qubit) is different from the classical bit in that its state is theoretically represented by a linear combination of two bases in the quantum state space (represented by a column vector of length 2). We can define two qubits |0$\rangle$ and |1$\rangle$, which can be described as follows: 
$$|0\rangle = \begin{bmatrix}1 \\0 \end{bmatrix} \hspace*{8mm} |1\rangle = \begin{bmatrix}0 \\1 \end{bmatrix}$$

\noindent
Qubits |0$\rangle$ and |1$\rangle$ are the computational basis state of the qubit. In other words, they are a set of basis of quantum state space.

Any qubit |$e\rangle$ can be expressed as a linear combination of two basis as $\alpha$|0$\rangle$ + $\beta$|1$\rangle$, where $\alpha$ and $\beta$ are complex numbers, and $|\alpha|^2+|\beta|^2 = 1$. This restriction is also called {\it normalization conditions}.

For example, 0.6|0$\rangle$ + 0.8|1$\rangle$ is a qubit state, and $\frac{1-i}{2}|0\rangle + \frac{1+i}{2}|1\rangle$ is also a qubit state, but (1-$i$)|0$\rangle$ + (1+$i$)|1$\rangle$ is not a legitimate qubit state because it does not satisfy the normalization condition. 

Intuitively, qubits can be viewed as a superposition of classical bits. For example, 0.6|0$\rangle$ + 0.8|1$\rangle$ can be considered as a superposition state of |0$\rangle$ and |1$\rangle$, where the probability for bit |0$\rangle$ is $0.6^2 = 0.36$ and for bit |1$\rangle$ is $0.8^2 = 0.64$. Note that $0.6^2 + 0.8^2 = 1$ .

\subsection{Quantum Gate}
\label{subsec:q-gate}

Just as a logic gate in a digital circuit that can modify the state of a bit, a quantum gate can change the state of a qubit. A quantum gate can have only one input and one output (transition of a single quantum state), or it can have multiple inputs and multiple outputs (transition of multiple quantum states). The number of inputs and outputs should be equal because the operators need to be reversible which means no information can be lost in quantum computing. Here, we describe two quantum gates with single input and output, and one quantum gate with multiple inputs and outputs.

\subsubsection{\textbf{NOT Gate}}
\label{subsubsec:not}

The NOT gate works on a single qubit. It can exchange the coefficients of two basis vectors:
$$NOT(\alpha |0\rangle + \beta |1\rangle) = \alpha |1\rangle + \beta |0\rangle$$

\noindent
The quantum NOT gate is an extension of the NOT gate in classical digital circuits. 

A single input-output quantum gate can be represented by a $2\ \times\ 2$ matrix. The state of a quantum state after passing through the quantum gate is determined by the value of the quantum state vector left-multiplied by the quantum gate matrix. The quantum gate matrix corresponding to the NOT gate is
$$X = \begin{bmatrix}0&1\\1&0\end{bmatrix}$$ 

\noindent
Therefore, the result of a qubit passing a NOT gate is

$$X \begin{bmatrix}\alpha \\ \beta \end{bmatrix} = \begin{bmatrix}0&1\\1&0\end{bmatrix} \begin{bmatrix}\alpha \\ \beta \end{bmatrix} = \begin{bmatrix}\beta \\ \alpha \end{bmatrix}$$

\begin{figure*}[h!] 
  \centerline{
\Qcircuit @C=0.8em @R=0.75em { 
  \lstick{\ket{j_{1}}}   &   \gate{H}  &   \gate{R_{2}}  &   \gate{R_{3}}  &   \qw & \cdots &        &  \gate{R_n}   &   \qw       &   \qw         &   \qw &   \qw    &  \qw   &  \qw            &  \qw      &   \qw  &  \qw    & \qw   & \qw            &   \qw        &   \rstick{\ket{y_1}} \qw        \\
  \lstick{\ket{j_{2}}}   &   \qw       &   \ctrl{-1}     &   \qw           &   \qw & \qw    &  \qw   &  \qw          &   \gate{H}  &   \gate{R_2}  &   \qw &   \cdots &        &  \gate{R_{n-1}} &  \qw      &   \qw  &  \qw    & \qw   & \qw            &   \qw        &   \rstick{\ket{y_2}} \qw       \\
  \lstick{\ket{j_{3}}}   &   \qw       &   \qw           &   \ctrl{-2}     &   \qw & \qw    &  \qw   &  \qw          &   \qw       &   \ctrl{-1}   &   \qw &   \qw    &   \qw  &  \qw            &  \gate{H} &   \qw  &  \cdots &       & \gate{R_{n-2}} &   \qw        &   \rstick{\ket{y_3}} \qw   \\
  \lstick{\vdots }       &             &                 &                 &       & \ddots &        &               &             &               &       &   \ddots &        &                 &           &        &  \ddots &       &                &              &   \rstick{\vdots }         \\
  \lstick{\ket{j_{n}}}   &   \qw       &   \qw           &   \qw           &   \qw & \qw    &  \qw   &  \ctrl{-4}    &   \qw       &   \qw         &   \qw &   \qw    &  \qw   &  \ctrl{-3}      &  \qw      &   \qw  &  \qw    & \qw   & \ctrl{-2}      &   \gate{H}   &   \rstick{\ket{y_{n}}} \qw
}
}
\caption{Quantum circuit for the quantum Fourier transform (QFT) algorithm.}
\label{figure:QFT_circuit}
\end{figure*}
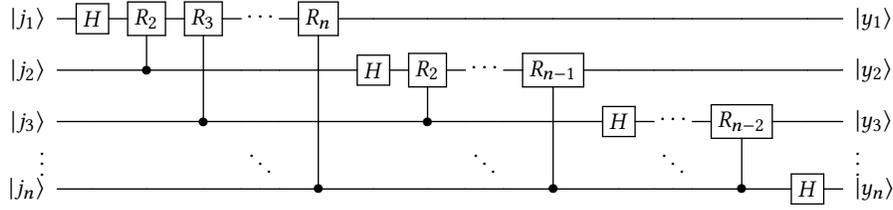

\subsubsection{\textbf{Hadamard Gate}}
\label{subsubsec:hadamard}

The Hadamard gate also works on a single qubit, which can decompose existing quantum states according to its coefficients:
$$H(\alpha |0\rangle + \beta |1\rangle) = \frac{\alpha + \beta}{\sqrt{2}}|0\rangle + \frac{\alpha - \beta}{\sqrt{2}}|1\rangle$$ 
This can be represented by a matrix as: 
$$H = \frac{\sqrt{2}}{2}\begin{bmatrix}1&1\\1&-1\end{bmatrix}$$
Although Hadamard gate is not directly related to the AND and OR gates in classical digital circuits, it has important applications in many quantum computing algorithms. Interested readers can try to prove that after applying the Hadamard gate twice in a row, the quantum state will return to its original state. This behavior is consistent with the NOT gate.

There can be an infinite variety of single input and output quantum gates, as long as the result of the left multiplication of the qubit state vector by the qubit state vector-matrix still satisfies the qubit normalization conditions. 

\subsubsection{\textbf{Controlled NOT Gate}}
\label{subsubsec:controlled}

Computer programs are full of conditional judgment statements: if so, what to do, otherwise, do something else. In quantum computing, we also expect that the state of one qubit can be changed by another qubit, which requires a quantum gate with multiple inputs and outputs. The following is the controlled-NOT gate (CNOT gate). It has two inputs and two outputs. If the input and output are taken as a whole, this state can be expressed by 
$$\alpha |00\rangle + \beta |01\rangle + \gamma |10\rangle + \theta |11\rangle$$ 
where $|00\rangle$, $|01\rangle$, $|10\rangle$, $|11\rangle$ are column vectors of length 4, which can be generated by concatenating $|0\rangle$ and $|1\rangle$. This state also needs to satisfy the normalization conditions, that is 
$$|\alpha|^2 + |\beta|^2 + |\gamma|^2 + |\theta|^2 = 1$$

The CNOT gate is two-qubit operation, where the first qubit is usually referred to as the control qubit and the second qubit as the target qubit. When the control qubit is in state |0$\rangle$, it leaves the target qubit unchanged, and when the control qubit is in state |1$\rangle$, it leaves the control qubit unchanged and performs a Pauli-X gate on the target qubit. It can be expressed in mathematical formulas as follows:
$$CNOT(\alpha |00\rangle + \beta |01\rangle + \gamma |10\rangle + \theta |11\rangle) = \alpha |00\rangle + \beta |01\rangle + \gamma |11\rangle + \theta |10\rangle$$
\noindent
The action of the CNOT gate can be represented by the following matrix:
$$X = \begin{bmatrix}1&0&0&0\\0&1&0&0\\0&0&0&1\\0&0&1&0\end{bmatrix}$$ 

\begin{figure*}[t]
\centerline{\includegraphics[width=0.8\linewidth]{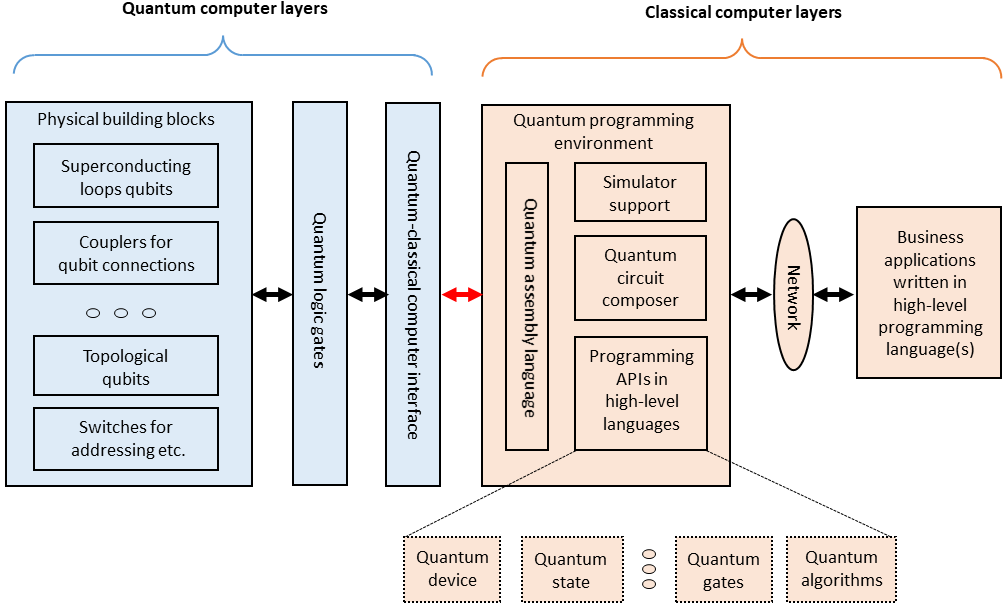}}
\caption{The architecture of quantum computing system in~\cite{sodhi2018quality}.}
\label{fig:architecture}
\end{figure*}

\subsection{Quantum Circuit}
\label{subsec:q-circuit}

Quantum circuits, also known as quantum logic circuits, are the most commonly used general-purpose quantum computing models, which represent circuits that operate on qubits under an abstract concept. 
A quantum circuit is a collection of interconnected quantum gates. The actual structure of quantum circuits, the number and the type of gates, as well as the interconnection scheme, are all determined by the unitary transformation $U$, performed by the circuit. The result of a quantum circuit can be read out through quantum measurements. 

As an example, Figure~\ref{figure:QFT_circuit} shows the quantum circuit for the quantum Fourier transform (QFT) algorithm. The quantum gates used in the circuit are the Hadamard gate and the controlled phase gate $R_{m} = \begin{bmatrix}1 & 0\\0 & e^{\frac{2{\pi}i}{2^{m}}}\end{bmatrix}$. The black dots present the control bits. 

\subsection{Superposition and Entanglement}
\label{subsec:q-entanglement}
Quantum computers use the laws of quantum mechanics to provide a computation mechanism that is significantly different from classical machines.
The first distinguishing feature of a quantum system is known as {\it superposition}~\cite{dirac1981principles,zeilinger1999experiment}, or, more formally, the superposition principle of quantum mechanics. A quantum system is actually in all its possible states at a time, rather than existing in one distinct state at the same time. For a quantum computer, this means that a quantum register exists in the superposition of all its possible 0's and 1's configurations, which is different from a classical system whose register contains only one value at any given time. Until the system is observed, it collapses into an observable, deterministic classical state. 

Quantum systems may also exhibit {\it entanglement}~\cite{einstein1935can,schrodinger1935discussion}, which is a quantum mechanical phenomenon. A state is considered to be entangled if it cannot be broken down into its more basic parts. In other words, two distinct elements of the system are entangled if one part of a system cannot be described without considering the other. A particularly interesting feature of quantum entanglement is that elements of a quantum system may be entangled even if they are separated by considerable space. Thus, measurements performed on one system can instantaneously influence other systems entangled with it. Quantum entanglement has applications in quantum computation~\cite{shor1999polynomial,nielsen2002quantum}, quantum cryptography~\cite{bennett2014quantum}, and quantum teleportation~\cite{gottesman1999quantum,jin2010experimental}.

\subsection{Quantum Measurement}
\label{subsec:q-measurement}

From the introduction of the quantum gate above, we can see that a qubit is the superposition state of two quantum states |0$\rangle$ and |1$\rangle$; two qubits combined into a whole are in a superposition of four quantum states |00$\rangle$, |01$\rangle$, |10$\rangle$, and |11$\rangle$. From this analogy, $n$ qubits can be described by a superimposed state of $2^n$ qubits, which is a huge advantage over $n$ classical bits that have only one fixed state. However, the laws of physics also have their limitations-in order to know the exact state of a qubit, a measurement is needed. But the measurement can cause the superposition state to collapse into a deterministic state. The information contained in a qubit after measurement is similar to a classical bit, and its value can only be 0 or 1. A superposed state can collapse to 0 or 1, the probability of which is determined by the coefficients $\alpha$ and $\beta$ in the superposed state. The probability that this superposed state collapses to |0$\rangle$ or |1$\rangle$ is $|\alpha|^2$ and $|\beta|^2$, respectively. Similarly, for a two-qubit system, the probability that the measurement results in |00$\rangle$, |01$\rangle$, |10$\rangle$, |11$\rangle$ is ${|\alpha|}^2$, ${|\beta|}^2$, ${|\gamma|}^2$, and ${|\theta|}^2$, respectively.

Such physical law causes the result of quantum computing to be non-deterministic. In actual applications, additional means are needed to verify the correctness of the output. For the algorithm used in quantum computing, if we can reduce the superposition state of qubits as much as possible in the previous step of measurement, we can get correct results with a high probability.

\subsection{Quantum Algorithm}
\label{subsec:q-algorithm}

Quantum algorithms are designed to solve classical problems in a probabilistic manner~\cite{montanaro2016quantum}. For example, the phase estimation algorithm (QPE)~\cite{kitaev2002classical} takes a matrix $U$, an eigenvector $|\psi\rangle$, as inputs, and calculates the eigenvector value of $U$ corresponding to $|\psi\rangle$. 
Another example is Grover's search algorithm~\cite{grover1996fast}: given a sparse non-zero function $f: \{0,\ .\ .\ .\ ,2^{n-1}\} \rightarrow \{0,\ 1\}$, Grover's algorithm outputs a value $x$ such that the probability of $f(x) = 1$ is high enough to beat brute force search, which is true in general.
Quantum algorithms are usually designed to solve (classical) problems more efficiently than existing classical algorithms.

A quantum algorithm is generally realized on a quantum circuit designed according to the parameters of the problem (such as the size of the instance), usually through the following three steps of iteration: 
\begin{itemize}
    \item[(1)]{\it Memory initialization}: preparing the initial state (basis state) $|x\rangle$ with $x \in \mathbb{B}^{n}$ (classical register),
    \item[(2)]{\it Operation of the quantum circuit}: applying a quantum circuit of polynomially many in $n$ gates from some universal gate set, and
    \item[(3)]{\it Performing an elementary measurement}: measuring the quantum state to retrieve classic data.
\end{itemize}

Quantum circuits are regarded as a predictive tool that can provide some (classical) information probabilistically, from which the target result can be inferred. The fact that the probability is high enough is a direct consequence of the mathematical properties of the unitary mapping described by the quantum circuit. The essence of quantum algorithms (and the reason for their efficiency) is to describe an effective circuit that implements this unitary mapping. Therefore, it is crucial to describe an efficient circuit that can realize the unitary mapping in order to obtain the efficiency of the quantum algorithm.

\subsection{Quantum Computing}
\label{subsec:q-computing}

The necessary procedure of quantum computing is as follows:

\begin{itemize}
  \item Start with a set of calculated ground states (each qubit is initialized to $|0\rangle$ or $|1\rangle$ as input to the calculation);
  \item Pass the qubits through a series of quantum gates according to a predetermined algorithm;
  \item A series of bits is obtained as a result of quantum measurement.
\end{itemize}

Note that the realization of each quantum gate requires manual manipulation of quantum gates.


\begin{table*}[htbp] %
\begin{threeparttable}
\centering
\caption{A Brief and Historical Summary of Quantum Programming Languages}
\label{table:QPLsummary}
\renewcommand\arraystretch{1.0}
\small
\begin{tabular}{|c|c|c|c|c|c|} 
\hline 
{\bf Year} & {\bf Language} & {\bf Reference(s)} & {\bf Semantics} & {\bf Host Language} & {\bf Paradigm} \\
\hline  
1996 & Quantum Lambda Calculi & \cite{maymin1996extending} & Denotational & lambda Calculus & Functional \\\hline
1998 & QCL & \cite{omer1998procedural,omer2000quantum,omer2003structured,omer2005classical} & & C & Imperative \\\hline
2000 & qGCL &\cite{sanders2000quantum,zuliani2001quantum,zuliani2001formal,zuliani2004non} & Operational & Pascal & Imperative \\\hline
2003 & $\lambda_{q}$ & \cite{van2003quantum,van2004lambda} & Operational & Lambda Calculus & Functional \\\hline  
2003 & Q language & \cite{bettelli2002architecture,bettelli2003toward} & & C++  & Imperative \\\hline  
2004 & QFC (QPL) & \cite{selinger2004towards,selinger2004towards+,selinger2006lambda} & Denotational & Flowchart syntax (Textual syntax) & Functional  \\\hline
2005 & QPAlg & \cite{jorrand2004quantum,lalire2004process} & & Process calculus & Other \\\hline
2005 & QML & \cite{altenkirch2005functional,altenkirch2005qml,grattage2006qml} & Denotational & Syntax similar to Haskell & Functional \\\hline
2004 & CQP & \cite{gay2004communicating,gay2005communicating,gay2006quantum} & Operational & Process calculus & Other \\\hline
2005 & cQPL &\cite{mauerer2005semantics} & Denotational & & Functional \\\hline
2006 & LanQ & \cite{mlnarik2006introduction,mlnarik2007quantum,mlnarik2007operational,mlnavrik2008semantics} & Operational & C & Imperative \\\hline
2008 & NDQJava & \cite{xu2008quantum} & & Java & Imperative \\\hline
2009 & Cove & \cite{purkeypile2009cove} & & C\# & Imperative \\\hline
2011 & QuECT & \cite{chakraborty2011quect} & & Java & Circuit \\\hline
2012 & Scaffold & \cite{abhari2012scaffold,javadiabhari2015scaffcc} & & C (C++)&  Imperative \\\hline
2013 & QuaFL & \cite{lapets2013quafl}& & Haskell &  Functional \\\hline
2013 & Quipper & \cite{green2013introduction,green2013quipper} & Operational & Haskell & Functional \\\hline
2013 & Chisel-Q & \cite{liu2013chisel} & & Scala & Imperative, functional \\\hline
2014 & LIQUi|$\rangle$ & \cite{wecker2014liqui} & Denotational & F\# & Functional \\\hline
2015 & Proto-Quipper & \cite{ross2015algebraic,rios2017categorical} & & Haskell & Functional \\\hline
2016 & QASM & \cite{pakin2016quantum} & & Assembly language & Imperative \\\hline
2016 & FJQuantum & \cite{feitosa2016fjquantum} & & Feather-weight Java & Imperative \\\hline
2016 & ProjectQ & \cite{haner2016high,projectq2017projectq,steiger2018projectq} & & Python & Imperative, functional \\\hline
2016 & pyQuil (Quil) & \cite{smith2016practical} & & Python & Imperative \\\hline
2017 & Forest & \cite{smith2016practical,regetti2017forest} & & Python & Declarative \\\hline
2017 & OpenQASM & \cite{cross2017open} & & Assembly language & Imperative \\\hline
2017 & qPCF &\cite{paolini2017mathsf,paolini2019qpcf}& & Lambda calculus & Functional \\\hline
2017 & QWIRE &\cite{paykin2017qwire} & Denotational & Coq proof assistant & Circuit \\\hline
2017 & cQASM & \cite{khammassi2018cqasm} & & Assembly language & Imperative \\\hline
2017 & Qiskit & \cite{ibm2017qiskit,gadi_aleksandrowicz_2019_2562111} & & Python & Imperative, functional \\\hline
2018 & IQu & \cite{paolini2019quantum} & & Idealized Algol & Imperative \\\hline
2018 & Strawberry Fields & \cite{killoran2018strawberry,killoran2019strawberry} & & Python & Imperative, functional \\\hline
2018 & Blackbird & \cite{killoran2018strawberry,killoran2019strawberry} & & Python & Imperative, functional \\\hline
2018 & QuantumOptics.jl & \cite{kramer2018quantumoptics} & & Julia & Imperative \\\hline
2018 & Cirq & \cite{cirq2018google} & & Python & Imperative, functional \\\hline
2018 & Q\# & \cite{svore2018q} & & C\# & Imperative \\\hline
2018 & $Q|SI\rangle$ & \cite{liu2018q} & & .Net language & Imperative \\\hline
2020 & Silq & \cite{bichsel2020sliq} & Operational & & Imperative, functional \\\hline
2020 & Quingo & \cite{team2020quingo} &  & Python & Imperative \\\hline
\end{tabular}

\begin{tablenotes}
    \item[$\ast$]{\bf Year}: The invented year of the language.
    \item[$\ast$]{\bf Language}: The name of the language.
    \item[$\ast$]{\bf Reference(s)}: The main reference paper(s) of the language.
    \item[$\ast$]{\bf Semantics}: The type(s) of semantics for the language the authors described in the reference(s).
    \item[$\ast$]{\bf Host language}: The classical language on (or to) which the language is based (or extended).
    \item[$\ast$]{\bf Paradigm}: We consider each language to belong to three types of paradigms: {\it imperative language}, {\it functional language}, and {\it circuit design language}.
\end{tablenotes}  
\end{threeparttable}
\end{table*}


In~\cite{sodhi2018quality}, a general architecture of a quantum computing system has been proposed. As shown in Figure~\ref{fig:architecture}, the architecture comprises of two parts: {\it quantum layer} and {\it classical layer}. The quantum layer contains purely quantum hardware and circuitry, and can be considered as comprising the quantum processing unit (QPU). The detailed composition of this layer is listed as follows.

\begin{itemize}
    \item[$\bullet$] {\it Physical building blocks} include quantum hardware that typically makes use of superconducting loops for the physical realization of qubits, and the physical qubit coupler/interconnect circuitry and other elements that are needed for qubit addressing and control operations.
    \item[$\bullet$] {\it Quantum logic gates} contain physical circuitry that makes up quantum logic gates.
    \item[$\bullet$] {\it Quantum-classical interface} includes the hardware and software which provides the interfacing between classical computers and a QPU.
\end{itemize}

The classical layer consists of classical hardware and software, as shown in the following:

\begin{itemize}
    \item[$\bullet$] {\it Quantum programming environment} provides the quantum assembly language that is necessary for instructing a QPU, the programming abstractions for writing quantum programs in a high-level programming language, and the simulator support, as well as IDEs.
    \item[$\bullet$] {\it Business applications} include quantum software applications that are written to cater to business requirements.
\end{itemize}


\section{Quantum Software Engineering}
\label{sec:QSD}

Quantum computing is not only a technological advancement, but can also be considered as a new general-purpose paradigm for software development, which can radically influence the way a software system is conceived and developed. This calls for new, quantum-specific, software engineering approaches. In this section, we first introduce quantum programming and define quantum software engineering. This will be followed by an introduction of a quantum software life cycle for quantum software development, which will be a basis for discussing the state of the art of quantum software engineering activities in the rest of the paper. 

\subsection{Quantum Programming}
\label{subsec:q-programming}

Quantum programming is the process of designing and building executable quantum computer programs to achieve a particular computing result~\cite{miszczak2012high,ying2016foundations}. Since the quantum programming efforts predate the other quantum software development techniques, we will focus first on quantum programming. Here, we briefly introduce the concepts, languages, and semantics of quantum programming. An excellent book written by Ying~\cite{ying2016foundations} covers much more similar materials to this section, with a focus on the fundamentals of quantum programming.  

\subsubsection{\textbf{Concepts of Quantum Programming}}
\label{subsubsec:concept}

A quantum program consists of blocks of code, each of which contains classical and quantum components. Quantum operations can be divided into {\it unitary} operations (reversible and preserve the norm of the operands), and {\it non-unitary} operations (not reversible and have probabilistic implementations). A quantum program executed on a quantum computer uses a quantum register of qubits to perform quantum operations, and a classical register of classic bits to record the measurements of the qubits' states and apply quantum operators conditionally~\cite{cross2017open}. Therefore, a typical quantum program usually consists of two types of instructions. One is called {\it classical instructions} that operate on the state of classical bits and apply conditional statements. Another is called {\it quantum instructions} that operate on the state of qubits and measure the qubit values.

\subsubsection{\textbf{Languages for Quantum Programming}}
\label{subsubsec:QPL-qpl}

Early quantum programming language development efforts focused on exploring the quantum Turing Machine (QTM) model proposed by Deutsch~\cite{deutsch1985quantum}, but did not result in practical tools for programming quantum computers. This situation made the quantum circuit models quickly become the driving force for quantum programming. To build it as a practical language (rather than just designing circuits), Knill~\cite{knill1996conventions,knill2000encyclopedia} proposed a pseudocode notion for quantum programming and the model of a quantum random-access machine (QRAM) in which the quantum system is controlled by a classical computer. This model influenced the design of subsequent quantum programming languages.
{\"O}mer~\cite{omer1998procedural,omer2000quantum,omer2003structured,omer2005classical} developed the first practical quantum programming language QCL with a C-like syntax in 1998. 
Since then, many quantum programming languages have been designed and implemented in terms of different types of language paradigms for programming quantum computers, including qGCL~\cite{sanders2000quantum,zuliani2001quantum,zuliani2001formal,zuliani2004non}, LanQ~\cite{mlnarik2006introduction,mlnarik2007quantum}, Scaffold~\cite{abhari2012scaffold,javadiabhari2015scaffcc}, Q language~\cite{bettelli2002architecture,bettelli2003toward}, NDQJava~\cite{xu2008quantum}, Q\#~\cite{svore2018q}, $Q|SI\rangle$~\cite{liu2018q}, ProjectQ~\cite{haner2016high,steiger2018projectq}, and Qiskit~\cite{gadi_aleksandrowicz_2019_2562111} for imperative quantum programming languages. Quantum lambda calculi~\cite{maymin1996extending}, QFC (QPL)~\cite{selinger2004towards,selinger2004brief,selinger2006lambda}, QML~\cite{altenkirch2005functional,altenkirch2005qml,grattage2006qml}, cQPL~\cite{mauerer2005semantics}, ~\cite{vizzotto2005concurrent}, QuaFL~\cite{lapets2013quafl}, Quipper~\cite{green2013introduction,green2013quipper}, LIQUi|>~\cite{wecker2014liqui}, qPCF~\cite{paolini2017mathsf, paolini2019qpcf}, and IQu~\cite{paolini2019quantum} for functional quantum programming languages, and QPAlg~\cite{jorrand2004quantum}, qASM~\cite{pakin2016quantum}, QuECT~\cite{chakraborty2011quect}, QWire~\cite{paykin2017qwire}, Quil~\cite{smith2016practical} for other quantum programming languages paradigms.  Table~\ref{table:QPLsummary} gives a summary of quantum programming languages since it first emerged in 1996. In the table, we summarize each quantum programming language according to six categories: {\it year}, {\it language}, {\it reference}, {\it semantics}, {\it host language}, and {\it language paradigm}. For more information regarding quantum programming languages, one can also refer to survey papers discussed in Section~\ref{sec:work}.

\subsubsection{\textbf{Semantics for Quantum Programming}}
\label{subsubsec:QPL-semantics}

Semantics for quantum programming have also been studied extensively, and recently many approaches~\cite{selinger2004towards,brunet2004dynamic,feng2005semantics,mauerer2005semantics,mlnarik2007operational, mlnavrik2008semantics,gay2010semantic,ying2012quantum,pagani2014applying,cho2016semantics,ying2016foundations,hasuo2017semantics,clairambault2019game}  have been proposed for describing the semantics of quantum programming languages. These approaches can be classified into three categories: {\it operational semantics}, {\it denotational semantics}, and {\it axiomatic semantics}. Although there is no survey for the semantics of quantum programming languages, readers can also refer to some survey papers~\cite{selinger2004brief,gay2005bibliography,jorrand2007programmer,ying2012quantum} and book~\cite{ying2016foundations} on quantum programming languages for the detailed discussions of the quantum language semantics issues.  

\subsection{Definition of Quantum Software Engineering}
\label{subsec:QSE-definition}

While quantum programming languages are exciting developments, coding is not the primary source of problems in quantum software development. Requirements and design problems are much more common and costly to correct. Therefore, the focus on quantum software development techniques should not be limited to quantum programming issues, but should also focus on other aspects of quantum software engineering. The promise quantum software development methodologies hold for overcoming complexity during analysis and design and accomplishing analysis and design reuse is really significant. If it is accepted that quantum software development is more than quantum programming, then a whole new approach, including life cycle, must be adopted to address other aspects of quantum software development.

Software engineering is a problem-solving process with roots in behavioral, engineering, project management, computer science, programming, cost management, and mathematical science. According to~\cite{fuggetta2000software}, the software engineering process can be defined as follows: 

\vspace*{2mm}
\noindent
{\it “A software process can be defined as the coherent set of policies, organizational structures, technologies, procedures, and artifacts that are needed to conceive, develop, deploy, and maintain a software product.”}
\vspace*{1mm}

For the definition of software engineering, the IEEE has developed a `standard' definition~\cite{ieee1990ieee} for classical software engineering as follows: 

\vspace*{2mm}
\noindent
{\it “(1) The application of a systematic, disciplined, quantifiable approach to the development, operation, and maintenance of software; that is, the application of engineering to software. (2) The study of approaches as in (1).”}
\vspace*{2mm}

Fritz Bauer~\cite{naur1969software} first stated the definition of software engineering as follows:

\vspace*{2mm}
\noindent
{\it “The establishment and use of sound engineering principles in order to obtain economically software that is reliable and works efficiently on real machines.”}
\vspace*{2mm}

\noindent
Although many definitions of software engineering~\cite{boehm1976software,andriole1993software,ieee1990ieee,finkelsteiin2000software,sommerville2011software,webster2020software} have been developed since then, Bauer's definition is still the widely accepted one~\cite{pressman2010software}, and can serve as the basis for our definition of quantum software engineering in this paper. 

In this paper, we define quantum software to include not only a set of executable quantum programs but also associated supporting libraries and documents needed to develop, operate, and maintain quantum software. By defining quantum software in this broader perspective, we hope to emphasize the necessity of considering timely documentation as an integral part of the quantum software development process. Inspired by the definitions of classical software engineering, we define quantum software engineering as following: 

\vspace*{2mm}
\noindent
{\it “Quantum software engineering is the use of sound engineering principles for the development, operation, and maintenance of quantum software and the associated document to obtain economically quantum software that is reliable and works efficiently on quantum computers.”}

\vspace*{2mm}

\noindent
In this definition, we would like to highlight three important issues in quantum software engineering. First, it is important to apply the “{\it sound engineering principles}” to quantum software development. Second, the quantum software should be built “{\it economically}.” Finally, the quantum software should be “{\it reliable}” and needs to work “{\it efficiently}” on quantum computers. 

\subsection{Quantum Software Engineering Methods, Tools, and Processes}
\label{subsec:QSE-process}

Quantum software engineering, as its classical counterpart~\cite{dorfman1997software}, can also be considered as a layered technology, which contains three elements: {\it methods}, {\it tools}, and {\it processes}. 

Quantum software engineering {\it methods} provide the techniques for constructing the quantum software. They consist of a wide range of tasks, including the design of data structures, program architecture, algorithm procedure, coding, testing, and maintenance. Quantum software engineering {\it tools} provide automated or semi-automated support for these methods. Quantum software engineering {\it processes} are the foundation for the quantum software engineering. The process provides the glue that holds the methods and tools together and enables the rational and timely development of quantum software. They define the sequence in which methods would be applied: the deliverables, the controls that help quality assurance and change coordination, as well as the milestones that enable quantum software managers to access the progress. 

Different ways of combining these three elements of quantum software engineering would lead to different quantum software engineering models. The selection of a suitable model should be based on the nature of the project and the application, the methods and tools to be used, and the controls and deliverables that are required. Three typical examples discussed in~\cite{pressman2010software} for classical software engineering are the classical life cycle (or waterfall model), the prototyping model, and the evolutionary model. Each of these models could be extended to the domain of quantum computing for supporting the development of quantum software systems. In Section~\ref{subsec:QSLC}, we will introduce a life cycle for quantum software, which is inspired by the idea of the waterfall model from classical software engineering.  

\begin{figure*}[t]
\centerline{\includegraphics[width=0.75\linewidth]{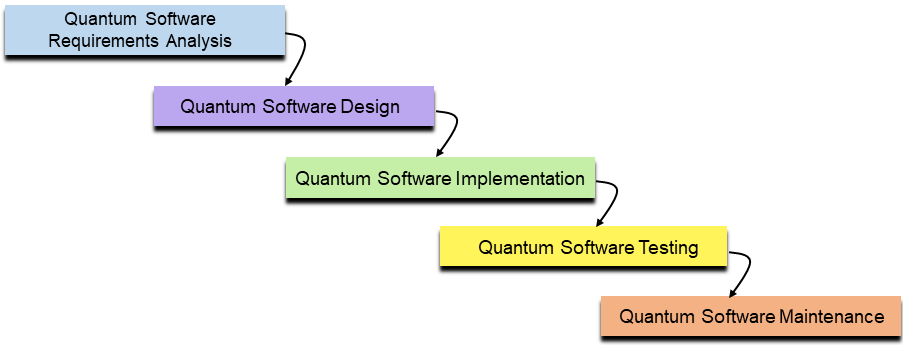}}
\caption{A quantum software life cycle.}
\label{fig:c-lifecycle}
\end{figure*}

\subsection{A Generic View of Quantum Software Engineering}
\label{subsec:QSE-view}

Pressman~\cite{pressman2010software} introduced a generic view of classic software engineering, which can be extended to the field of quantum software engineering. Quantum software engineering can be regarded as a branch of systems engineering~\cite{everitt2016quantum,finkelsteiin2000software,chestnut1967systems}, which involves the development of large and complex quantum software systems. To this end, the following issues must be discussed.

\begin{itemize}
    \item What is the problem to be solved by quantum software? 
    \item What characteristics of the quantum software are used to solve the problem?
    \item How will the quantum software (and the solution) be implemented?
    \item What method will be used to detect errors made in the design and construction of the quantum software?
    \item How will the quantum software be supported in a long period, when the users request corrections, adaptations, and enhancements?
\end{itemize}

To adequately engineering quantum software, a quantum software engineering process must be defined. In this section, we consider the generic characteristics of the quantum software process. The work related to quantum software engineering can be categorized into three generic phases: {\it definition}, {\it development}, and {\it maintenance} phases. These phases are independent of the application domain, project size, or complexity.

The definition phase focuses on {\it what}, dealing with the problems that quantum software developers try to determine. These issues include topics such as what information to process, what functions and performance are required, what interfaces to build, what design constraints exist, and what verification standards are needed to define a successful quantum software system. At this stage, there may be three sub-processes, including quantum software system analysis, quantum software project planning, and quantum software requirements analysis.

The development phase focuses on {\it how}, which deals with the problems that quantum software developers try to describe. These issues include how to design quantum software architecture and related data structures, how to implement quantum program details, how to translate quantum software design into a quantum programming language, and how to conduct quantum software testing. This phase may also contain some sub-processes, including quantum software design, quantum software implementation, and quantum software testing.

The maintenance phase focuses on {\it change}, which is related to error correction, adaptions required as the quantum software environment evolves, and modifications due to enhancements brought about by changes in customer requirements. The maintenance phase reapplies the definition and development phases, but it is carried out in the context of existing quantum software.

\begin{table*}[h]
\centering
\caption{Brief Summary of the Patterns for Quantum Algorithms in~\cite{leymann2019towards}}
\label{table:pattern}
\renewcommand\arraystretch{1.1}
\small
\begin{tabular}{|l|p{9cm}|} 
\hline 
\multicolumn{1}{|c|}{\bf Pattern Type} & \multicolumn{1}{c|}{\bf Description}  \\
\hline  
(1) Initialization (aka state preparation) & Initializing the input of a quantum register in a straightforward manner \\
\hline
(2) Uniform superposition & \tabincell{l}{Creating an equally weighted superposition of all possible states of the qubits\\ of a quantum register}  \\
\hline
(3) Creating entanglement & Creating an entangled state \\
\hline
(4) Function table & Computing a function table of a finite Boolean function \\  
\hline
(5) Oracle (aka black box) & Reusing the computation of another quantum algorithm \\
\hline
(6) Uncompute (aka unentangling aka copy-uncompute) & Removing entanglement that resulted from a computation  \\
\hline 
(7) Phase shift & Distinguishing important aspects of a state efficiently \\
\hline
(8) Amplitude amplification & Increasing the probability of finding a solution \\
\hline
(9) Speedup via verifying & Achieving a speedup when verifying a solution is simple  \\
\hline 
(10) Quantum-classic split & Splitting a solution between a quantum computer and a classical computer \\
\hline
\end{tabular}
\end{table*}

\subsection{\bf Quantum Software Life Cycle}
\label{subsec:QSLC}

A software life cycle model can be defined as a reference model for a software development process that supports the design and construction of high-quality software~\cite{ghezzi2002fundamentals}. The software life cycle provides a well-structured phased process that can help organizations quickly produce high-quality software, which has been well tested and ready for production use.
Several software life cycle models have been proposed for classical software development, including the waterfall model~\cite{royce1970managing}, evolutionary model~\cite{hirsch1985evolutionary}, and the Spiral model~\cite{boehm1988spiral}. Among them, the most widely accepted life cycle model for classical software development is the waterfall life cycle model, which is sometimes called the classical model. Other models are often improved upon it. 

As the first step, this paper introduces a systemic, sequential life cycle model for the quantum software development process to support the design and construction of quantum software systems. A quantum software life cycle model begins at the system level and progresses through requirements analysis, design, implementation, testing, and maintenance. The model, however, is extensible in the sense that one can add more phases into it if necessary. Figure~\ref{fig:c-lifecycle} shows the life cycle, which encompasses the following phases, each one is flowing into the next.

\begin{itemize}
    \item {\it Quantum software requirements analysis}
    \item {\it Quantum software design}
    \item {\it Quantum software implementation}
    \item {\it Quantum software testing}
    \item {\it Quantum software maintenance}
\end{itemize} 

In the following, we briefly introduce each phase of the life cycle model from the perspective of quantum software development.

\subsubsection{\textbf{Quantum Software Requirements Analysis}} 

The quantum software life cycle model begins with the requirements analysis phase, where the stakeholders discuss the requirements of the software that needs to be developed to achieve a goal. The requirements analysis phase aims to capture the detail of each requirement and to make sure everyone understands the scope of the work and how each requirement is going to be fulfilled. The analysis creates a set of measurable quantum software requirements that specify, from the supplier's perspective, what characteristics, attributes, and functional and performance requirements the quantum software system is to possess, to satisfy stakeholder requirements. Later life cycle phases, including design, implementation, testing, maintenance for quantum software, assume that requirements analysis continues through the quantum software life cycle. 

\subsubsection{\textbf{Quantum Software Design}}
Quantum software design is the second phase in the life cycle, which is a process to transform user requirements into a suitable form that helps the programmer in quantum software implementation (coding). 

As classical software design~\cite{dorfman1997software}, quantum software design may also involve two stages: {\it architectural design} and {\it detailed design}. {\it Architectural design} defines a collection of quantum software components, their functions, and their interactions (interfaces) to establish a framework for the development of a quantum software system. {\it Detailed design} refines and expends the architectural design to describe the internals of the quantum software components (the algorithms, processing logic, data structures, and data definitions). Detailed design is complete when the description is sufficient for implementation (coding).  

\subsubsection{\textbf{Quantum Software Implementation}} 
After completing the requirements and design activities, the next phase of the life cycle is the implementation or development of quantum software. At this phase, developers start coding based on the requirements and design discussed in the previous phases. Developers also perform the unit testing for each component to test the new code they write, review the code for each other, create builds, and deploy quantum software to the environment. This development cycle is repeated until the requirements are met.

\subsubsection{\textbf{Quantum Software Testing}}
Testing is the last phase of the quantum software life cycle before the software is delivered to customers. During testing, testers start to test the quantum software according to the requirements. The aim is to find defects within the software as well as to verify whether the software behaves as expected based on the documentation from the quantum software requirements analysis phase.

\subsubsection{\textbf{Quantum Software Maintenance}}
Quantum software maintenance is the last phase of the quantum software life cycle, which represents all the modifications and updates made after the delivery of quantum software products. 


\section{Quantum Software Requirements Analysis}
\label{sec:requirement}

To the best of our knowledge, until recently, no research work has been proposed to address the issues on quantum software requirements analysis. However, as quantum software development resources are going to be accumulated, we believe that the development of methodologies, techniques, and tools to support quantum software requirements analysis will become a critical and inevitable issue.

\section{Quantum Software Design}
\label{sec:design}

Developing quantum algorithms is considered extremely difficult comparing to classical counterparts due to some intrinsic features of quantum computing, such as superposition and entanglement. Hence, new design principles for quantum algorithms are strongly demanded. Although software design is recognized as a critical phase in classical software engineering, researches~\cite{Perez-Delgado2020quantum,cartiere2016quantum,thompson2018quantum} on the principle and methodology for quantum software design are just starting. This section gives an overview of state of the art on quantum software design.  

\subsection{Quantum Software Modelling}
\label{subsec:d-modelling}

\subsubsection{\textbf{UML-Based Modelling language}}
The Unified Modeling Language (UML)~\cite{boochunified,rurnbaughunified,jacobson1999unified} is a general-purpose, well-known modeling language in the field of classical software engineering. It provides a standard way to visualize the design of the classical software development life cycle. It seems reasonable to extend the UML approach to support quantum system design.

P\'{e}rez-Delgado and Perez-Gonzalez~\cite{Perez-Delgado2020quantum} presented an approach to extending the UML to model quantum software systems. The extension covers two types of UML diagrams: {\it class diagram} and {\it sequence diagram}. Extension for the other three components (i.e., composite structure diagrams, activity diagrams, and state machine diagrams) still needs to be studied. They also shared three fundamental observations from their work, including:

\begin{itemize}
    \item The nature of quantum computation, leading to the intrinsic difference between quantum and classical computation is in how it achieves its goal,
    \item Quantum computation changes the very nature of information itself, and is much more productive and powerful than classical computation, and
    \item The classical vs. quantum nature of the information used by a module is an important consideration when discussing about its internal implementation and interface.
\end{itemize}

Their observations, though not complete, can serve as a base for further studying the design methodologies of quantum software systems.  

\subsubsection{\textbf{Generic Modelling Languages}}
Unlike the previously mentioned approaches, Ali and Yue ~\cite{ali2020modeling} presented a preliminary conceptual model of quantum software from the perspective of model-based engineering. Their model includes key concepts of quantum software such as quantum variables, quantum states, and quantum operations and is independent of any modeling languages, quantum programming languages, and quantum computing platforms. In this way, the proposed model could be used as a basis for the development of new modeling languages for quantum software, which is either as a new domain-specific language (DSL) or as an extension of existing modeling languages, such as the UML as described in~\cite{Perez-Delgado2020quantum}. As a case study, they presented the preliminary results of using the conceptual model to model the state-based behavior of a quantum program that implements quantum entanglement. They also showed some research directions and open issues about quantum software modeling and analysis.

\subsection{Quantum Software Specification}
\label{subsec:d-specification}
Quantum computing relies on quantum mechanics, which is a subject more familiar to physicists rather than computer scientists and software engineers. Thus, we must be aware of the underlying theory before we reason about quantum computers. Even though delivering the principles of quantum mechanics is not an easy task, due to their counter-intuitive nature. 

Recently, the method for reasoning about quantum computing is a mixture of linear algebra and Dirac representation, which is a subject more suitable for physicists than computer scientists (software engineers)~\cite{cartiere2016quantum}. Therefore, it is necessary to provide a more "intuitive" way to think and write quantum algorithms, thereby simplifying the design and implementation of quantum software. This can be achieved, for example, by introducing a specification language, which adopts the symbolism and reasoning principle of software engineering and will take on the role of Hilbert space algebra, allowing us to describe quantum structures and design quantum algorithms in a more natural way~\cite{grattage2006qml}.

To this end, Cartiere~\cite{carmelo2013quantum,cartiere2016quantum} presented some work on defining a formal specification language for quantum algorithms. The goal is to introduce a formal specification language that can be used to represent the basic notations of quantum computing, which are more suitable for quantum computing scientists and quantum software engineers. The language is based on Z~\cite{abrial1980specification,woodcock1996using}, a well-studied formal specification language for classical software systems. The language can represent some elementary quantum gates such as {\it Identity gate}, {\it Pauli-X gate}, {\it Phase Shift gate}, {\it Pauli-Z gate}, {\it Hadamard gate}, and {\it C-NOT gate}, which perform unitary transformations to support writing quantum programs. Cartiere also showed how to use the language to specify some simple quantum algorithms, such as the Deutsch-Jozsa algorithm~\cite{deutsch1985quantum,deutsch1992rapid}. It is, however, unclear if the language can be used to represent more complex quantum algorithms such as Shor's {\it integer factoring} quantum algorithm~\cite{shor1999polynomial} and Grover's {\it database search} quantum algorithm~\cite{grover1996fast}. 

\subsection{\bf Pattern Language for Quantum Software}
\label{subsec:q-pattern}

In classical computing, software engineers have long noticed that specific patterns recur and persist across different applications and systems. A pattern is a systematic design that can capture the experiences of experts on good designs or best practices, and record the essence of this wisdom in an easy-to-understand way for peers to use~\cite{alur2003core}. Patterns can be used to describe and reason about the relationship between a specific context, and a problem and its solutions in that context. A pattern language is a network of related patterns that may jointly contribute to an encompassing solution of a complex problem~\cite{alexander1977pattern,beck1987using}. Patterns and pattern languages are used in multiple design and engineering disciplines, especially for software architecture design~\cite{frank1996pattern,buschmann2007pattern,fowler2002patterns,volter2006software}. 

Leymann~\cite{leymann2019towards} proposed a pattern language for supporting the development of quantum algorithms. There is a need for documenting solutions for problems recurred in quantum software development, as observed and mentioned in~\cite{nielsen2002quantum,lipton2014quantum}, which informally summarized some basic "tricks" used in quantum algorithms. The main contribution of this work is to systematize this to become a subject of a software engineering discipline for quantum algorithms. 
In~\cite{leymann2019towards}, Leymann identified ten types of basic patterns derived from existing quantum algorithms that mainly based on gate models. Each pattern is described through eight elements, including {\it name}, {\it intend}, {\it icon}, {\it problem statement}, {\it context}, {\it solution}, {\it know uses}, and {\it next}. The summary of these ten types of patterns is listed in Table~\ref{table:pattern}. 

Additionally, Leymann also discussed some issues on the using of these patterns regarding software engineering, and mentioned that the pattern language proposed can be stored as a pattern repository. In the repository, each pattern could be linked with the corresponding implementation in a concrete quantum programming language to support the programming of the pattern. Note that the patterns identified in~\cite{leymann2019towards} is in no way meant to be exhaustive, and more patterns should be identified in the future to make the patterns and the pattern language practical useful.  

\subsection{Modular Design of Quantum Systems} 
\label{subsec:modular-design}

As quantum computation becomes more and more complex, the modular design holds significant practical value. Just as classical computation is modular, quantum computing also needs modular design methodology. 

Thompson {\it et al.}~\cite{thompson2018quantum} presented a formal framework to specify modularity in quantum systems to fill this gap. Their research can be summarized in two parts. First, they formalized the modularity of quantum computing within a server-client framework. Such modularity ensures that a client can construct a $U$-independent device that implements a $U$-dependent quantum operation $P[U]$ by blindly invoking the server to implement a unitary $U$.
They also proposed the necessary constraints on such modular architectures and explored their implications for existing quantum algorithms, including quantum factorization~\cite{shor1994algorithms} and DQC1~\cite{knill1998power}, which operate by probing the properties of some input-dependent unitary operator $U$.
The resulting "modularity constraint" suggests that the above algorithms are typically non-modular. So it is impossible to prefabricate devices that are valid for every possible input - forcing their circuits to be tailored for each specific input. Second, they explored ways to circumvent this constraint. They figured out what functions needed to be sacrificed to refine the existing algorithm to restore the modular implementation. This led to two new algorithms.

\begin{itemize}
\item A universal device that can evaluate the normalized modulus of the trace of any exponentially large unitary $U$, even if this unitary is entirely unknown and provided in a black box.
\item A modular factorization algorithm that can factorize the numbers while recycling a more significant part of its circuit structure, thus significantly reducing the number of control gates required for the implementation.
\end{itemize}

Although this work is more focused on the modular circuit design of quantum computing rather than modular software system design, both the ideas are similar. The presented framework could also be adapted to support the development of modular quantum software systems.

Sánchez and Alonso~\cite{sanchez2021definition} discussed the concept of module, a key concept in the software engineering discipline, and tried to establish the initial criteria for determining the cohesion and coupling levels of a module in quantum programming in order to build a sound quantum software engineering. 

\section{Quantum Software Implementation}
\label{sec:implementation}

This section will be brief, as much of the material has been covered in the survey (or overview) papers of quantum programming languages~\cite{selinger2004brief,gay2006quantum,sofge2008survey,miszczak2011models,ying2012quantum,valiron2013quantum,valiron2015programming,ying2016foundations,chong2017programming,spivsiak2017quantum,garhwal2019quantum,zorzi2019quantum}.

The initial focus of quantum software development on the implementation level has resulted in a range of quantum programming approaches. It is crucial to identify a suitable quantum programming technique when implementing a quantum application. Most 
quantum programming techniques support one or more base programming languages. For example, languages are available for quantum programming in C (QCL~\cite{omer1998procedural}), C++ (Scafflod~\cite{abhari2012scaffold}), C\# (Q\#~\cite{svore2018q}), Python (ProjectQ~\cite{steiger2018projectq}, Qiskit~\cite{ibm2017qiskit}, Forest~\cite{smith2016practical}),  F\# (LIQUi|$\rangle$~\cite{wecker2014liqui}), Scala (Chisel-Q~\cite{liu2013chisel}), and Haskell (Quiper~\cite{green2013quipper}). Therefore, the first natural step in the choice of a suitable technique is to reduce the available set of techniques to those that support the base programming language to be employed in application development. If no suitable technique is available for the base programming language, a custom language needs to be implemented. 
%

\begin{table*}[ht]
\centering

\caption{Bug types and their corresponding defense types for quantum software in~\cite{huang2018qdb,huang2019statistical}}
\label{table:bugtype}
\renewcommand\arraystretch{1.1}
\small
\begin{tabular}{|l|l|} 
\hline 
\multicolumn{1}{|c|}{\bf Bug Type} & \multicolumn{1}{c|}{\bf Defense Strategy}  \\
\hline  
(1) Incorrect quantum initial values & Assertion checks for classical and superposition preconditions \\
\hline
(2) Incorrect operations and transformations & Assertion checks for unit testing\\
\hline
(3) Incorrect composition of operations using iteration & Assertion checks for classical intermediate states \\
\hline
(4) Incorrect composition of operations using recursion & Assertion checks for entangled intermediate states\\  
\hline
(5) Incorrect composition of operations using mirroring & Assertion checks for product state postconditions\\
\hline
(6) Incorrect classical input parameters & Assertion checks for classical postconditions  \\
\hline 
(7) Incorrect deallocation of qubits & Assertions on algorithm postconditions\\
\hline
\end{tabular}
\end{table*}

\section{Quantum Software Testing}
\label{sec:testing}

Quantum computers are powerful. However, since the human intuition is much better adapted to the classical world than the quantum world, programmers might make mistakes more frequently when designing programs on quantum computers compared to programming on classical counterparts \cite{ying2012floyd}. Furthermore, compared to the classical computers, quantum computers have very different properties, such as quantum superposition, entanglement, and no-cloning~\cite{nielsen2002quantum}. Therefore, the prediction of the behavior of quantum software is difficult~\cite{chong2017programming,ying2012floyd}. As a result, new quantum software testing and debugging techniques are needed to discover~\cite{stepney2006journeys}. Recently, researches are emerging for identifying bugs, and for testing and debugging of quantum software. This section gives an overview of state of the art on quantum software testing, in a broader sense, according to quantum software {\it bug type}, {\it assertion}, {\it testing}, {\it debugging}, and {\it analysis}. 

\subsection{Bugs in Quantum Software} 
\label{subsec:bugs}

A software bug is regarded as the abnormal program behaviors which deviate from its specification~\cite{allen2002bug}. Bug patterns are erroneous code idioms or bad coding practices that have been proved to fail repeatedly, usually caused by the misunderstanding of a programming language’s features, the use of erroneous design patterns, or simple mistakes sharing common behaviors. 
To debugging and testing quantum software, it is essential to deeply understanding the behavior of bugs in quantum programs.  

\subsubsection{\textbf{Bug Types and Patterns}}
\label{subsubsec:bug-pattern}
Huang and Martonosi~\cite{huang2018qdb,huang2019statistical} studied the bug types for special quantum programs to support quantum software debugging. Based on the experiences of implementing some quantum algorithms, they identified several types of bugs specific to quantum software and proposed some defense strategies for each type of bug. These bug types include incorrect quantum initial values, incorrect operations and transformations, the incorrect composition of operations using iteration, the incorrect composition of operations using recursion, incorrect composition of operations using mirroring, incorrect classical input parameters, and incorrect deallocation of qubits. They also proposed some defense strategies for each type of bug. 
A detailed summary of these bugs and their corresponding strategies can be found in Table~\ref{table:bugtype}.  

Zhao {\it et al.}~\cite{zhao2021identifying} identified and categorized some bug patterns in the quantum programming language Qiskit to provide both researchers and programmers a clear view of what types of bugs may happen in quantum programs and how to detect them. Their study of bug patterns mainly focuses on bug pattern symptoms, root causes, and cures and preventions. For each bug pattern, they also provide an example to illustrate the pattern’s symptoms. Identifying such patterns in a quantum programming language can help programmers improve their productivity in finding bugs and reduce software maintenance costs. Their research can be regarded as the first step to provide an underlying basis for debugging and testing quantum programs. The research is primary, which does not use every possible quantum-related construct or cover all characteristics of a quantum programming language. Therefore, new research should be carried out to cover other remaining quantum-related constructs and their interactions.

\subsubsection{\textbf{Bug Benchmarks}}
\label{subsubsec:bug-benchmark}
Campos and Souto~\cite{campos2021qbugs} proposed QBugs, which is a collection of reproducible bugs in quantum algorithms for supporting controlled experiments for quantum software debugging and testing. QBugs proposes some initial ideas on building a benchmark for providing an experimental infrastructure to support the evaluation and comparison of new research and the reproducibility of published research results on quantum software engineering. It also discusses some challenges and opportunities on the development of QBugs. 

Zhao {\it et al.}~\cite{zhao2021bugs4q} proposed Bugs4Q, a benchmark of thirty-six real, manually validated Qiskit bugs from four popular Qiskit elements (\texttt{Terra}, \texttt{Aer}, \texttt{Ignis}, and \texttt{Aqua}), supplemented with the test cases for reproducing buggy behaviors. Bugs4Q collects reproducible bugs in Qiskit programs and supports downloading and running test cases for quantum software testing. Each actual bug and the corresponding fixes are publicly available for research. Bugs4Q collects almost all the existing bugs of Qiskit on GitHub and updates them in real-time. Furthermore, these programs are sorted separately and filtered except for the bugs with originally available test cases and support for reproduction. Bugs4Q provides a database that includes an analysis of bug types to classify existing bugs for experimental evaluation of isolated bugs.

\subsection{Assertions for Quantum Software} 
\label{subsec:q-assertion}

An assertion is a statement about the expected behavior of a software component that must be verified during execution~\cite{foster2006assertion}. In software development, a programmer defines an assertion to ensure a specific program state at run time. Assertions have been used extensively for detecting runtime faults, documenting programmer intent, and formally reasoning about the correctness of classical programs~\cite{clarke2006historical}. Recently, several researches have been carried out for defining and identifying assertions in quantum software systems.

\subsubsection{\textbf{Invariant and Inductive Assertions}}
\label{subsubsec:invariant}

Ying {\it et al.}~\cite{ying2017invariants} studied the issue of how to define the notion of invariant and inductive assertion~\cite{floyd1993assigning} for quantum programs. They considered this issue in two different ways -- from the perspectives of additive and multiplication. They proved that both additive and multiplication invariants could be used to prove the partial correctness of quantum programs. They also showed that additive invariants could be derived from additively inductive assertions, and can be generated through an SDP (Semidefinite Programming~\cite{vandenberghe1996semidefinite}) solver. However, how to generate the multiplication invariants is still an open problem that has not been addressed in~\cite{ying2017invariants}, and therefore still needs to be explored.

\subsubsection{\textbf{Applied Quantum Hoare Logic}}
\label{subsubsec:aQHL}

Hoare logic (also known as Floyd-Hoare logic)~\cite{hoare1969axiomatic, floyd1967assigning} is a formalism with a set of logic rules used for rigorous reasoning on the correctness of computer programs. Hoare logic has been extensively used for the verification, debugging, and testing of classical software~\cite{adrion1982validation,apt2019fifty}. Recently, researchers have extended Hoare logic to the quantum domain to support the formal verification of quantum programs~\cite{brunet2004dynamic,baltag2006lqp,d2006quantum,kakutani2009logic,ying2012floyd,unruh2019quantum,zhou2019applied}. Among them, Li {\it et al.}~\cite{zhou2019applied} introduced applied quantum Hoare logic (aQHL), which is a simplified version of quantum Hoare logic (QHL)~\cite{ying2012floyd}, with particular emphasis on supporting debugging and testing of the quantum programs. aQHL simplified two issues of QHL through: (1) binding QHL to a particular class of pre- and post-conditions (assertions), that is, projections, to reduce the complexity of quantum program verification and to provide a convenient way used for debugging and testing of quantum software. (2) providing additional rules for reasoning about the robustness of quantum programs. As aQHL can be used to specify assertions such as pre- and post-conditions, and invariants for quantum software, hopefully, it could provide a very efficient way to support assertion-based quantum software debugging and testing. Although no detail ~\cite{zhou2019applied} was given on how aQHL could be used to debug and test quantum software, we believe it could be a promising way for tracking errors in quantum software, similar to its classical counterpart. Moreover, besides aQHL, similar approaches that use other types of (dynamic) quantum logic~\cite{brunet2004dynamic,baltag2006lqp,kakutani2009logic} for specifying assertions for supporting debugging of quantum software should also be investigated. 
\subsubsection{\textbf{Assertion Library for Quantum Software}}
\label{subsubsec:property}

As we will discuss in Section~\ref{subsec:q-testing}, Honarvar {\it et al.}~\cite{honarvar2020property} proposed a property-based approach to testing quantum programs in Q\# (developed by Microsoft)~\cite{svore2018q}. To this end, they developed a testing specification language consisting of a library of assertion methods that can be used to specify various properties (assertions) in Q\# programs for testing. These assertion methods include \verb+AssertProbability+, \verb+AssertEntangled+, \verb+AssertTeleported+, \verb+AssertTransformed+, and \verb+AssertEqual+. \verb+AssertProbability+ uses a robust statistical method to test the probability of which observes a qubit in a given state. \verb+AssertEntangled+ takes two qubits as its arguments to test whether or not they are entangled. \verb+AssertEqual+ tests the equality of two states in quantum programs. \verb+AssertTeleported+ tests quantum teleportation as it is a significant protocol in the quantum realm. \verb+AssertTransformed+ tests the validity of any unitary transformation. The detailed description of each assertion method is summarized in Table~\ref{table:AssertionMethod}. Recently, the properties in the specification language are stateless, which talk about the input-output relations without carrying much of the past state of the program. However, it is possible to enrich the property specification language to allow for stateful properties where the history of the past operations may be used in the property that is to be checked.
Moreover, although those assertion methods developed by Honarvar {\it et al.}~\cite{honarvar2020property} are used initially for testing, they also have the potential to be used in quantum program debugging and runtime error checking.  

\begin{table*}[ht]
\centering
\caption{Brief Summary and Comparison of Assertion Methods in~\cite{honarvar2020property} and Assertion Functions in~\cite{liu2020quantum}}
\label{table:AssertionMethod}
\renewcommand\arraystretch{1.2}
\footnotesize
\begin{tabular}{|l|p{9cm}|} 
\hline 
\multicolumn{1}{|c|}{\bf Assertion Methods~\cite{honarvar2020property}} & \multicolumn{1}{c|}{\bf Description}  \\
\hline  
\verb+AssertProbability(q0,state,probability)+ & To take three arguments to test the expected \verb+probability+ of observing a qubit \verb+q0+ in a given \verb+state+ after measurement. \\
\hline
\verb+AssertEntangled(q0,q1)+ & To take two qubits \verb+q0+ and \verb+q1+ as its arguments to test whether or not they are entangled. \\
\hline
\verb+AssertEqual(q0,q1)+ & To take two qubits \verb+q0+ and \verb+q1+ as its arguments to compare the equality of these two states. \\
\hline
\verb+AssertTeleported(q0,q1)+ & To take the sent and received qubits \verb+q0+, \verb+q1+ as its arguments to test quantum teleportation.\\  
\hline
\verb+AssertTransformed(q0,(+$\theta$ \verb+interval)(+$\phi$ \verb+interval))+ & To test the validity of any unitary transformation.\\
\hline
\hline 
\multicolumn{1}{|c|}{\bf Assertion Functions~\cite{liu2020quantum}} & \multicolumn{1}{c|}{\bf Description}  \\
\hline  
\verb+classical_assertion(circuit,qubitList,value)+ & To take three arguments to specify the quantum circuit under test, the list of qubits for assertion, and a particular classical value to assert for. \\
\hline
\verb+entanglement_assertion(circuit,qubitList,flag)+ & To take three arguments specifying the quantum circuit under test, the list of qubits for assertion, and the type of entanglement. \\
\hline
\verb+superposition_assertion(circuit,qubitList,phaseDict,flag)+ & To take four arguments specifying the quantum circuit under test, the list of qubits for assertion, the quantum state dictionary for the qubits, and a flag. \\
\hline
\end{tabular}
\end{table*}

\subsection{Quantum Software Testing}  
\label{subsec:q-testing}

Testing~\cite{myers1979art,beiser1983software} is the process that executes a program with the intent to find errors, which is a very critical process for supporting quality assurance during software development. In this section, we review state of the art for testing quantum software.

\subsubsection{\textbf{Challenges on Quantum Software Testing}}
\label{subsubsec:open-problem}

Miranskyy and Zhang~\cite{miranskyy2019testing} proposed some challenges associated with white-box and black-box testings as well as verification and validation of quantum programs. They have shown that some of the existing software engineering practices are readily transferable to the quantum computing domain (e.g., code review). Others are difficult to transfer (e.g., interactive debugging). And the rest have to be introduced (e.g., complexity-dependent placement of a validation program). Rather than proposing a particular testing method (or strategy) for quantum software, they tried to define the software engineering practices for quantum software. With this definition at hand, software engineering specialists in academia and industry can start exploring this fascinating area of quantum computing and expand their work to other areas of testing as well as the remaining phases of the quantum software life cycle.

Usaola~\cite{usaolaquantum} proposed some ideas and identified some challenges when applying classical software testing to the quantum computing domain. He discussed how some prevalent strategies of classical software testing, such as {\it functional testing}, {\it white-box testing}, and {\it model-based testing}, can be applied to test quantum software. Functional testing~\cite{myers2011art} is a type of software testing that validates the software system against the functional requirements/specifications. Usaola discussed some basic concepts regarding functional testing (such as test case and test suite) and showed that a process of functional testing for quantum software might consist of the following three steps: 
\begin{enumerate}
    \item The initial situation of a quantum test case sets up the initial status of the qubits.
    \item Similar to classical testing, the quantum circuit is executed.
    \item The test suite saves the obtained result to calculate the most probable result.
\end{enumerate}

White-boxing testing~\cite{myers2011art} is a method that tests a software solution's internal structure, design, and coding. Among different white-box testing methods, Usaola specially studied mutation testing, which might be a good candidate technique for quantum software testing. He also gave a simple example program of adding two integer numbers using IBM's Qiskit simulator to show how the mutation testing can be applied for testing. However, how to define suitable quantum mutant operators to support quantum mutation testing has not been discussed. Model-based testing~\cite{utting2010practical} is a software testing technique where run time behavior of the software during the test is checked against predictions made by a model. To apply model-based testing to the quantum software, one should first use some modeling language, for example, UML, to model the behavior of the quantum software. Then, based on the model, one can generate test cases for the software under test to perform the testing. As we discussed in Section~\ref{sec:design}, one may use the Q-UML~\cite{Perez-Delgado2020quantum}, an extension of UML to the quantum domain, to model a quantum software system to support the test case generation of the system. 

\subsubsection{\textbf{Testing Approaches}}
\label{subsubsec:testing-approach}

There are many testing approaches proposed for classical software, however, few has been proposed for quantum software. 

\vspace*{1.5mm}
\noindent
{\textit{$\bullet$ Fuzz Testing}.}\hspace*{1.3mm} 
Fuzz testing (or fuzzing)~\cite{takanen2018fuzzing} is a software testing technique that inputs invalid or random data called {\it fuzz} into the software system to discover coding errors and security loopholes. 
Wang {\it et al.}~\cite{wang2018quanfuzz} adapted an existing technique from classical software testing called coverage-guided fuzzing (CGF)~\cite{zalewski2007american,craig2002systematic,serebryany2015libfuzzer} to test quantum software. They proposed \texttt{QuanFuzz}, a search-based test input generator for quantum software. The basic idea of \texttt{QuanFuzz} is to define some quantum sensitive information to evaluate the test inputs for quantum software and use a matrix generator to generate test cases with higher coverage. The testing process of \texttt{QuanFuzz} consists of two steps: 

\begin{enumerate}
    \item Extracts quantum sensitive information from the quantum source code. Such information includes the measurement operations on the quantum registers and the sensitive branches associated with the measurement results.
    
    \item Uses the sensitive information guided algorithm to mutate the initial input matrix and selects those matrices which improve the probability weight for a value of the quantum register to trigger the sensitive branch. 
\end{enumerate}
\noindent
This process keeps iterating until the sensitive branch is triggered. 
They also implemented \texttt{QuanFuzz} based on $Q|SI\rangle$~\cite{liu2018q} and {\it Nrefactory} for quantum software simulation and code instrumentation respectively, and evaluated \texttt{QuanFuzz} on seven quantum programs with different registers (containing 2 to 8 qubits), which are build-in benchmark programs from $Q|SI\rangle$. The experimental result showed that \texttt{QuanFuzz} can obtain 20\%$\sim$60\% more branch coverage than the classical test input generation method.    

\vspace*{1.5mm}
\noindent
{\textit{$\bullet$ Property-based Testing}.}\hspace*{1.3mm} 
Property-based testing~\cite{fink1997property,claessen2011quickcheck} uses specifications of essential properties to produce testing criteria and procedures which focus on these properties in a systematic manner. It has been shown to be a promising tool for generating test cases that reveal program faults in classical software. 

Honarvar {\it et al.}~\cite{honarvar2020property} presented \texttt{QSharpCheck}, a property-based testing approach for quantum software written in Q\#. To this end, they defined a testing specification language to specify the properties of Q\# programs as assertions. The whole design of the language is inspired by several ideas, such as the syntax of Q\#, quantum predicates and predicate transformers by D'Hondt and Panangaden~\cite{d2006quantum}, and quantum Hoare logic by Ying~\cite{ying2012floyd}. In the language, a test is represented by four parts: test property name and parameters, allocation and setup, function call, and assert and deallocate. They also identified several types of assertions that formed the basis of post-conditions and assertion types associated with the test specification language. Based on these specified properties, one can generate various types of test cases, run the test cases to test Q\# programs, and check the output results to see if there are any problems within the programs. Moreover, they also showed some basic considerations on the design and implementation of \texttt{QSharpCheck} tool and carried out two case studies to demonstrate the effectiveness of \texttt{QSharpCheck} through mutation analysis of Q\# programs. 

Li {\it et al.}~\cite{li2020projection} proposed {\it Proq}, a projection-based runtime assertion tool for testing and debugging quantum programs. {\it Proq} uses projection~\cite{birkhoff1936logic} to represent assertions, which, compared to classical representation, has more expressive power. Moreover, since projection naturally matches the projective measurement, which may not affect the measured state when the state is in one of its basis states~\cite{li2019poq}, it may reduce the testing overhead. 

\vspace*{1.5mm}
\noindent
{\textit{$\bullet$ Search-based Testing}.}\hspace*{1.3mm} 
Search-Based Software Testing (SBST)~\cite{mcminn2011search} is the use of optimizing search techniques, such as a Genetic Algorithms, to solve problems in software testing. Recently, SBST is showing promising in classical software testing, such as test case generation, test case prioritization, and test suite minimization.
Wang {\it et al.}~\cite{wang2021generating} proposed a search-based approach, called QuSBT (QUantum Search-Based Testing), to testing quantum programs. They aim to automatically generate test suites of a specific size based on the available test budget to maximize the number of failed test cases in the test suite. QuSBT includes the definition of problem codes, failure types, test evaluation with statistical tests, and the number of failed test cases in the test suite, fitness functions, and test case generation with genetic algorithms. To validate the effectiveness of QuSBT, six quantum programs were selected as benchmark programs, and 30 error versions were carefully designed to compare with random search (RS). The experimental results show that QuSBT provides a feasible solution for testing quantum programs.

\subsubsection{\textbf{Test Adequacy Criteria}}
\label{subsubsec:testing-criteria}
Test adequacy criteria are used to provide quantitative measurement on the degree of the target software that has been tested. Up to the present, many adequacy criteria are proposed and widely adopted in classical software testing, such as line coverage, branch coverage, data flow coverage. However, due to fundamental differences in programming paradigm and logic representation format for quantum software and traditional software, new test adequacy criteria are required to consider the characteristics of quantum software into consideration.

Ali \textit{et al.}~\cite{ali2021assessing} presented an approach \textit{Quito} to testing quantum software in a black-box way. \textit{Quito} defined three coverage criteria based on the inputs and outputs of a quantum program, as well as their test generation strategies. \textit{Quito} also defined two types of test oracles, as well as a method to determine the passing and failing of test suites with statistical analysis.
Let $Q$ be a quantum program, the set $V_{I}$ of \textit{valid input values} of $Q$ is the set of input values for that the program computation makes sense. The set $V_{O}$ of \textit{valid output values} of $Q$ is the set of output values that, according to the program specification, can be produced at least by an input value. Based on these, three coverage criteria can be defined as follows.

\begin{itemize}
\item The \textit{input coverage criterion} for $Q$ requires that for each valid input value $i \in V_{I}$, there exists a test $t = v$.
\item The \textit{output coverage criterion} for $Q$ requires that each valid output value $o \in V_{O}$ is observed at least once, i.e., there exists a test $t$ whose result is $Q(t) = o$.
\item The \textit{input-output coverage criterion} for $Q$ requires that, for each input-output pair $(i, o)$ where the probability of occurrence of output value $o$ for input value $i$ is not $0$, there exists a test $t = i$ whose result is $Q(t) = o$.
\end{itemize}

To evaluate the cost-effectiveness of the three coverage criteria, the experiments have been conducted with five quantum programs, which use mutation analysis to determine the coverage criteria’ effectiveness and cost in terms of the number of test cases. They also identified equivalent mutants for quantum programs based on the results of mutation analysis. 

Although testing is a widely used technique to guarantee the reliability of a software system, as Dijkstra stated in~\cite{dahl1972structured}, {\it testing can be used to show the presence of bugs, but never to show their absence}. Therefore, after testing, systematical techniques such as {\it debugging} are still needed to localize the bugs in the system.  

\subsection{Quantum Program Debugging}
\label{subsec:q-debugging}

 The process of finding and fixing bugs is called debugging, which is the activity of analyzing the program to locate and correct errors in a program by reasoning about causal relations between bugs and the error detected in the program~\cite{bradley1985science,bentley1985programming}. Program debugging tools are essential for any programming environment~\cite{araki1991general,lencevicius2000advanced}. In the life cycle of software development, almost 25\% of maintenance is carried out for debugging~\cite{lientz1980software}. Debugging methodologies and techniques are also crucial for quantum programming~\cite{chong2017programming}. This section gives an overview of quantum program debugging techniques from four aspects: {\it debugging tactics}, {\it debugging quantum processes}, {\it assertion-based debugging}, and {\it language support for debugging}. 

\subsubsection{\textbf{Debugging Tactics}} 
\label{subsubsec:debugging-tactic}

For classical program debugging, some well-established common tactics including {\it brute force}, {\it backtracking}, and {\it cause elimination}~\cite{myers2011art} are widely used. These debugging tactics can also be used to support the quantum program debugging. 

Miranskyy {\it et al.}~\cite{miranskyy2020your} analyzed the debugging tactics for quantum programs and discussed the possibility of applying some well-established classical debugging techniques to the quantum domain. They mainly considered three types of debugging tactics in classical debugging, as mentioned above, and concluded that {\it backtracking} tactics, especially those based on code reviews and inspections, could probably be applied for quantum debugging. This is also confirmed by the discussions with practical quantum software developers who usually use code reviews and inspections for their daily programming tasks~\cite{miranskyy2020your}. 

For {\it cause elimination} tactics, they pointed out that it could be naturally extended to the domain of quantum computing. First, one can set a hypothesis and specify a root cause for a bug understudy during quantum debugging, and then collect data and perform some experiments based on the data to prove the hypothesis. However, considering the probabilistic nature of the quantum program's behavior~\cite{nielsen2002quantum,miranskyy2019testing}, the final result of the quantum program should be assessed based on the distribution of the results obtained from multiple executions of the program. To solve the problem, Miranskyy {\it et al.}~\cite{miranskyy2020your} suggested to apply classical testing techniques of probabilistic programs~\cite{dutta2018testing,dutta2019storm} to the quantum computing domain.  

For {\it brute force} tactics, the most common tactic for classical debugging, its applicability to quantum debugging might depend on two situations. If we regard the quantum program as a black box, one could follow the classical debugging process: (1) First, trace the input and the output of the program (2) Then, record the input and output data in a log file. (3) Finally, analyze and compare these data against expected values to see if they are consistent. If we treat the quantum program as a white box, one has to consider how to capture the execution traces during the execution of the program to perform interactive debugging. However, considering the specific features such as superposition, entanglement, and no-cloning, in quantum computing, it is almost impossible. Miranskyy {\it et al.}~\cite{miranskyy2020your} discussed some scenarios regarding those specific quantum features for which the classical debugging cannot be applied, and suggested some potential solutions to solve these problems. Those solutions, however, might still be immature, thus may be infeasible in practice. Therefore, much work remains to be done to work out some solutions to support quantum program debugging.   

\subsubsection{\textbf{Assertion-based Debugging}} 
\label{subsubsec:assertion-debugging}
Several assertion-based approaches have been proposed recently for debugging quantum programs. As we described previously, Huang and martonosi~\cite{huang2018qdb,huang2019statistical} summarized several types of bugs in quantum programs based on their experiences of implementation of a set of quantum programs (or algorithms~\cite{shor1999polynomial,grover1996fast,olson2017quantum,mcardle2018quantum}). They also proposed some corresponding defense strategies in terms of programming and assertion to prevent such bugs during programming to develop bug-free quantum programs. This study is preliminary, and further, Huang and Martonosi~\cite{huang2019statistical} presented the statistical assertions for quantum programs based on statistical tests on some classical observations. These allow the programmers to decide if a quantum program state matches its expected value in one of classical, superposition, or entangled types of states. Based on this, they classified possible assertions in quantum software into three types: {\it classical assertion}, {\it superposition assertion}, and {\it entanglement assertion}. They extended an existing quantum programming language called Scaffold~\cite{abhari2012scaffold,javadiabhari2015scaffcc} with the ability to specify quantum assertions, and developed a tool to check these assertions in a quantum program simulator called QX~\cite{khammassi2017qx}. To demonstrate the effectiveness of their assertion-based approach, they performed three case studies on three benchmark quantum programs: factoring integers, database search, and quantum chemistry, each of which represents a different class of quantum algorithms. They also showed some results on what types of bugs are possible and laid out a strategy for using quantum programming patterns to place assertions and prevent bugs. Moreover, to validate the proposed approach, they cross-validated the quantum programs and the simulation results through the functional equivalent programs implemented in different quantum programming languages, including LIQUi|$\rangle$~\cite{roetteler2017design}, ProjectQ~\cite{haner2016high,steiger2018projectq}, and Q\#~\cite{svore2018q}. Through this validation, they found and shared some results and insights on how language features of different quantum programming languages can help insert the quantum assertions in a suitable place in the program, or otherwise prevent bugs in the first place.

Zhou and Byrd~\cite{zhou2019quantum} and Liu {\it et al.}~\cite{liu2020quantum} observed that the critical limitation of Huang and Martonosi's assertion-based approach~\cite{huang2019statistical} is that each measurement during debugging has to stop the execution of the program, and the assertions require aggregates of runs when the actual computation results are to be measured. To overcome this limitation, motivated by quantum error correction (QEC)~\cite{nielsen2002quantum,gottesman2010introduction} and nondestructive discrimination (NDD) ~\cite{jain2009secure}, they proposed an approach to constructing suitable quantum circuits to support runtime assertion check of quantum software~\cite{liu2020quantum,zhou2019quantum}. The key idea of this approach is to use ancilla qubits (some additional quantum bits) to collect the information of the qubits under test indirectly and to measure those ancilla qubits, rather than the original qubits under test directly. This can avoid disrupting the program execution during the assertion check at runtime. To this end, they identified three types of dynamic assertions, respectively, for classical values, entanglement, and superposition. {\it Assertion for classical value} is to ensure that the qubits are initialized to the correct value, or some intermediate classical results should satisfy some condition such as (|$\psi\rangle$==|0$\rangle$). {\it Assertion for entanglement} is to assert two or more qubits are in entangled states based on checking the parity of two qubits. {\it Assertion for superposition} is to assert that the use of Hadamard gates to set the input qubits in the uniform superposition state, and also to assert arbitrary superposition state represented as $|\psi\rangle = \sin(\frac{\theta}{2})|0\rangle + e^{i\varphi} cos(\frac{\theta}{2})|1\rangle$. 

The implementation of the idea of assertion circuits is also discussed. This is based on Qiskit developed by IBM~\cite{ibm2017qiskit}, an open-source framework for quantum computing, through integrating three types of assertion functions into the Qiskit development environment. Programmers can use these functions to instrument the necessary assertion circuits for classical, entanglement, and superposition states, and therefore can check the corresponding ancilla qubits for error detection. These three kinds of assertion functions include \verb+classical_assertion+, \verb+entanglement_assertion+, and \verb+superposition_assertion+. Table~\ref{table:AssertionMethod} summarises and compares these assertion functions, with the assertion methods proposed in~\cite{honarvar2020property}. The experimental results for several case studies have confirmed the effectiveness of the proposed approach on debugging as well as on improving the success rate for quantum algorithms such as Quantum Fourier Transform (QFT)~\cite{coppersmith2002approximate}, Quantum Phase Estimation (QPE)~\cite{kitaev2002classical}, and Bernstein-Vazirani algorithm~\cite{bernstein1993quantum,bernstein1997quantum}.

The assertion schemes proposed by Huang and Martonosi~\cite{huang2018qdb}, Zhou and Byrd~\cite{zhou2019quantum}, and Liu {\it et al.}~\cite{liu2020quantum}, however, still have some limitations that they can only handle limited quantum states related assertions and limited assertion locations, which may increase the difficulty in debugging based on the assertions. To overcome this problem, Li {\it et al.}~\cite{li2019poq,li2020projection} proposed {\it Proq}, a projection-based runtime assertion checker for debugging quantum programs. {\it Proq} uses projection~\cite{birkhoff1936logic} to represent assertions, which, compared to classical representation, has more expressive power. Moreover, since projection naturally matches the projective measurement, which may not affect the measured state when the state is in one of its basis states~\cite{li2019poq}, it may reduce the testing overhead. {\it Proq} can assert sophisticated quantum algorithms. To validate the approach, Li {\it et al.} performed both theoretical analysis and case studies. They performed two case studies on applying the projection-based assertions to two sophisticated quantum algorithms, the Shor’s algorithm~\cite{shor1999polynomial} and the HHL algorithm~\cite{harrow2009quantum}. Both theoretical analysis and case studies show that the proposed assertion-based approach is superior to the existing quantum program assertion~\cite{huang2019statistical,liu2020quantum}, with stronger expressive ability, more flexible assertion location, fewer execution times and lower implementation overhead.

\subsubsection{\textbf{Debugging Quantum Processes}}
\label{subsubsec:quantum-process}

Since the measurement of a quantum system may cause its state to collapse, it is challenging to debug quantum programs through the monitoring of the program. To solve this problem, Li and Ying~\cite{li2014debugging} proposed an approach to debugging quantum processes through monitoring measurements. The basic idea of the approach is to develop a protocol to support the debugging of quantum processes through monitoring, which does not disturb the states of the processes. 

Informally, the approach can be described as following. Suppose that we have built a quantum system to execute some quantum processes. The system is set to an initial state $|\psi\rangle$, and then evolve under the controlled Hamiltonian denoted by $H(t)$. This guarantees that the trajectory \{$|\psi\rangle$\} of the system states is anticipated. Since the time for the whole process is usually much longer than the time of a single quantum component (such as a gate) action, it may be considered as infinite. Here, we can see that if there is a bug of the system in the process at time $t'$, the system will not truly be $H(t)$ for $t \geq t'$ during the execution, which may cause errors in the system state, so we write $\rho_{t}$. For this situation, to debug the process, we should try to find a projection operator $P$ of the system as well as a sequence of time points $t_{1}, t_{2}, \dots (t_{n} \rightarrow \infty)$, such that $P|\psi_{t_{n}}\rangle=0$ for all $n$. Here, $P|\psi_{t}\rangle=0$ means that nothing can be detected by $P$ if the system state is $|\psi_{t}\rangle$ as anticipated. So, to perform debugging, we can monitor the process at time $t_{1}, t_{2}, \dots,$ by using a measurement apparatus formalized by $P$, which is called a {\it monitoring measurement}, and together with probability $tr(P_{\rho_{t_{n}}})$, the error state could be detected at time $t_{n}$. If this really happened, an error would be detected in the state, and $t'$ is more likely in $[t_{n-1}, t_{n}]$, and the relevant components should be carefully checked. Practically, the time points $t_{1}, t_{2}, \dots$ are determined by a classical program $S$. The critical point for this debugging approach is to find the required projector $P$, and the condition $P|\psi_{t}\rangle=0$ guarantees that the anticipated process is not disturbed by $P$. On the other hand, it is also implied that the debugging procedure is conclusive in the sense that if the process runs correctly, it would be no error reported. Besides this, the authors also gave a formal definition of the proposed debugging approach in the case of discrete-time evolution. 

As mentioned in the conclusion of the paper, the proposed debugging approach can only handle the debugging of the quantum processes with time-independent Hamiltonians, and the debugging with time-dependent hamiltonians is still an open problem~\cite{li2014debugging}.     

\subsubsection{\textbf{Language Support for Debugging}} 
\label{subsubsec:debugging-language}

The above work on tackling bugs in quantum programs via assertion checking is promising, but it usually has to check assertions dynamically during runtime, which could waste the quantum computing (or simulation) resources. To overcome this, inspired by previous work on the verification of quantum programs such as quantum weakest preconditions~\cite{d2006quantum} and quantum Floyd-Hoare logic~\cite{ying2012floyd}, Singhal and Reppy ~\cite{Singhal2020a,Singhal2020b,Singhal2020c} proposed to encode assertions into a static type system of a quantum programming language to support programmers to write correct programs from the start. In this way, it is hopeful that the programmers can encode some of the semantic properties that they intend for their programs as specifications in their code, so that a type checker could be used to ensure the correctness of some of those properties during compilation time. 
The basic idea, to this end, is to extend the Hoare Type presented in Hoare Type Theory (HTT) in~\cite{nanevski2008hoare} for classical programming languages, to the quantum Hoare Types in the quantum domain which can be encoded into a quantum programming language to support static checking, and even formal verification of quantum software.
A Hoare type, similar to the Hoare triples and is represented as $$\{P\} x: A \{Q\},$$ can be ascribed to a stateful computation if when executed in a heap satisfying the precondition {\it P}, the computation diverges or results in a heap satisfying the postcondition {\it Q} and returns a value of type {\it A}~\cite{nanevski2008hoare}. To support quantum circuits, the Hoare type is extended with its syntax augmented with that of QWire, a quantum circuit language that supports quantum circuit programming and verification. Through this way, the host language HTT can be augmented with a wire type $W$ in QWire, so that to treat quantum circuits as data, it can be used to specify properties in quantum programs as well. Here, $W$ and $A$ can be described as:
$$W =\ 1\ |\ bit\ |\ qbit\ |\ W1 \otimes W2$$ and $$A\ ::=\ . . .\ |\ Unit\ |\ Bool\ |\ A \times A\ |\ Circuit(W1, W2),$$ respectively.
This work aims to build a unified system eventually for supporting quantum programming, specification, and verification, but it is still preliminary recently, and more work remains to be done to realize the goal.

\subsection{Quantum Program Analysis}
\label{subsec:q-analysis}
Program analysis utilizes static techniques for reliable computation of the information about the dynamic behavior of programs~\cite{nielson2015principles}. Example applications include compilers (for code improvement) and software validation (for detecting errors). Several researches have been carried out for program analysis of quantum programs recently, which can support bug detection for quantum programs. 

\subsubsection{\textbf{Entanglement Analysis}}
\label{subsubsec:entanglement-analysis}
ScaffCC~\cite{javadiabhari2014scaffcc,javadiabhari2015scaffcc} is a scalable compiler framework developed at Princeton University for the quantum programming language Scaffold~\cite{abhari2012scaffold}. ScaffCC supports both compilation and compiling-time analysis for Scaffold programs to optimize and detect possible errors in the code. One analysis that ScaffCC supports is called {\it entanglement analysis}, which can conservatively identify each possible pair of qubits that might be entangled in the program. Such entangle information can help a programmer to design algorithms and to debug. To perform entanglement analysis, ScaffCC explores data-flow analysis techniques to automatically track the entanglements within the code through annotating the output of the QASM-HL program, an intermediate representation of ScaffCC, to denote possibly entangled qubits. The analysis is conservative in the sense that it assumes that if two qubits interact, they are likely to have become entangled with each other. However, this might lead to some false positive entangled qubit pairs when the number of the qubits is large. Moreover, entanglement analysis can also help to find the un-computing portions in the module, through analysis of un-entanglement, which can be created by applying inverse functions such as CNOT and Toffoli operations to the same set of control and target qubits.  

\subsubsection{\textbf{Robustness Analysis}}
\label{subsubsec:robustness-analysis}

Hung {\it et al.}~\cite{hung2019quantitative} presented a semantics for describing quantum computation with errors and an analysis that bounds the distance between the result of a noisy program and its corresponding ideal program on the same input. They used the analysis to compute error bounds for noisy versions of the quantum Bernoulli factory and quantum walk programs. They also showed how the analysis could be used to compute the error bounds for small circuits with and without error correction, showing examples of when using error correction is s beneficial and when there are trade-offs between the efficiency of error corrections and related costs.

\section{Quantum Software Maintenance}
\label{sec:maintenance}
 Support for software maintenance is crucially important in the development of computer software. A motivation for this is the well-known fact that somewhere between 60$\%$ and 80$\%$ of the cost of software comes from the maintenance phase of software development~\cite{vandoren1997maintenance}. The primary purpose of quantum software maintenance is to modify and update software applications after delivery to correct faults and to improve performance. Research work on this area~\cite{Castillo2020reengineering,kruger2020quantum} is emerging recently, which mainly focuses on the reengineering of existing classical software systems to integrate with new quantum algorithms.

\subsection{Reengineering Classical Information Systems to Quantum computing}
\label{subsec:reengineering}

Software Reengineering~\cite{chikofsky1990reverse} refers to the inspection and modification of the target software system with a series of activities such as design recovery, re-documentation, reverse engineering, and forward engineering. It aims to reconstruct the existing system as a new form to develop higher quality and more maintainable software.

Although significant progress has been made in the development of quantum computers, it is evident that quantum computers can not be used for all things in the short term (due to its high initial cost, among other reasons). Instead, it is common to use quantum computers to solve some important problems by making specific calls from classical computers to remote quantum computers in the cloud~\cite{stepney2006journeys}. In this case, most enterprises require to integrate and migrate their first quantum algorithm or future quantum algorithm with their existing classical information systems. Therefore, reengineering must be revisited to deal with the problems associated with the migrations of quantum computing and the next coexistence of classical and quantum software. 

P{\'{e}}rez{-}Castillo~\cite{Castillo2020reengineering} proposed a software modernization approach (model-driven reengineering)~\cite{seacord2003modernizing} to restructuring classical systems together with existing or new quantum algorithms to provide target systems that combine both classical and quantum information systems. The software modernization method in classical software engineering has been proved to be an effective mechanism that can realize the migration and evolution of software while retaining business knowledge. The solution proposed is systematic and based on existing, well-known standards such as Unified Modelling Language (UML)~\cite{boochunified,rurnbaughunified,jacobson1999unified} and Knowledge Discovery Metamodel (KDM)~\cite{perez2011knowledge}. The solution could benefit from reducing the development of new quantum information systems. Moreover, since it is based on international standards to represent the knowledge in an agnostic manner, it would be independent of any quantum programming languages. Besides, P{\'{e}}rez{-}Castillo also pointed out that quantum technologies and programming have not yet been addressed with techniques, good practices, and development methodologies of software engineering to meet quantum programs'needs.  

Another issue regarding reengineering (maintenance) is how to integrate quantum computation primitives (for instance, quantum software components) to an existing classical software system. Since quantum computer (QC) is very different from the previous technology and method, integrating QC into existing software systems requires that it not only solves this problem at the level of algorithm implementation, but also involves many more extensive problems studied in software engineering~\cite{pressman2010software}. 
Quantum annealing~\cite{kadowaki1998quantum,finnila1994quantum,shin2014quantum} is a quantum-mechanical metaheuristic (generic and approximate method) to solve combinatorial optimization and sampling problems. Kr{\"{u}}ger and Mauerer~\cite{kruger2020quantum} performed a case study on how to augment a quantum software component implemented in a quantum annealer (QA) for the Boolean satisfiability problem (SAT)~\cite{cook1997finding}, to an existing software system. In this case study, they discussed the related quality measures of quantum components, and showed that the method of transforming the SAT into a QA, which is mathematically equivalent but structurally different, can lead to substantial differences in these qualities. Their research also showed that defining and testing specific quality attributes of QC components, as studied in~\cite{sodhi2018quality}, is one of the key challenges. Although these properties do not play a core role in classical engineering, they must be considered in the software architecture with quantum components. Their study may help readers form a realistic intuition about the ability, potential, and challenges of using quantum components to enhance classical software in the near and medium term. They also claimed that in the current development stage, this problem must be considered at a much lower level than the conventional abstraction level in classical software engineering. The research also implies that the ability to easily and smoothly replace functional components of classical software architecture with quantum components, just like in classical component-based software engineering~\cite{heineman2001component}, is crucial for the success of the project.  

\section{Quantum Software Reuse}
\label{sec:reuse}

Computer software can be systematically reused across the entire development life-cycle, i.e., requirement specification, design, and implementation~\cite{krueger1992software,frakes2005software}. It has its place even in the post-delivery stages of development, e.g., its continuing quality assessment or software maintenance. This section gives an overview of state of the art on quantum software reuse in several aspects, including {\it quantum pattern reuse}, {\it quantum circuit reuse}, and {\it quantum state reuse}.

\subsection{Quantum Pattern Reuse}
\label{subsec:pattern-reuse}

As we discussed in Section~\ref{sec:design}, Leymann~\cite{leymann2019towards} identified some quantum patterns which can help develop quantum algorithms from the perspective of software reuse. They also plan to represent these quantum patterns in a pattern repository, which is a specialized database that stores pattern documents and manages the links between them. Such a quantum pattern repository allows a quantum software developer to query the database to find appropriate quantum patterns (e.g., to determine the entry pattern corresponding to a problem), supports browsing the content of each pattern document, and enables navigating between patterns based on the links between them. In this way, a quantum software developer could find suitable patterns from the repository, that may cross to several different domains, and compose them to solve a complex problem.  

\subsection{Quantum Circuit Reuse}
\label{subsec:circuit-reuse}

The design of efficient quantum circuits is an essential problem in quantum computation~\cite{williams1998automated,perkowski2003hierarchical}. For a given unitary matrix, it is a difficult task to find a highly optimized quantum circuit. There are a few known methods of quantum circuit design, which are mainly based on heuristic search techniques such as genetic algorithm and simulated annealing. However, these methods are limited to a relatively small circuit size, and the solutions generated by the methods are usually difficult to explain. To solve these problems, Klappenecker and R\"{o}tteler~\cite{klappenecker2003quantum} presented a new design principle for quantum circuits that is based exactly on the idea of reusing some existing quantum circuits in the construction of other quantum circuits. The unique characteristics of the design principle is: assuming that we have an effective set of quantum circuits available, we can systematically construct efficient quantum circuits by reusing and combining a group of highly optimized quantum circuits. The sound mathematical basis of this design method allows meaningful and natural explanations of the generated circuits. They also suggested that from a practical perspective, it would be interesting to build a database of medium-sized matrix groups with effective quantum circuits. A given transformation can then be searched in this database by linear algebra, and automatically deriving quantum circuit implementations in this way could be an attractive possibility. 

Allouche {\it et al.}~\cite{allouche2017reuse} attempted to extend further the design-by-reuse method proposed by Klappenecker and R\"{o}tteler~\cite{klappenecker2003quantum} to a general framework to support circuit synthesis through reusing the existing quantum circuits. In their extension, the approach needs to find suitable groups for the implementation of new quantum circuits. They also identified some critical points which are necessary for constructing their extended method. For example, this method relies on the distribution of the information between the group and the coefficient matrix. When the group contains enough information, such as the Fourier transform power group, the coefficient matrix is easy to calculate, and the efficiency of the quantum Fourier transform synthesis is used to generate an effective circuit on non-trivial operators. Besides, they also studied some potential group candidates, such as the {\it projective Pauli group} and the {\it dihedral group}, for testing the proposed method.

\subsection{Quantum State Reuse}
\label{subsec:state-reuse}

“{\it Some quantum state are hard to create and maintain, but are a valuable resource for computing. Twenty-first-century entrepreneurs could make a fortune selling disposable quantum states}~\cite{preskill1999plug}.”

To make it real, Preskill presented an interesting idea called {\it plug-in quantum software} for probably reusing quantum states, which make up a quantum software program~\cite{preskill1999plug}. The basic idea is that manufacturers can design a valuable quantum state and use a special-purpose device to make multiple copies of that quantum state; these copies can be tested to ensure their quality and stored until they are needed. Consumers can pay to download that state and plug it into their quantum computer for a performance boost. 

A candidate application of plug-in quantum software is to ensure the reliable operation of quantum computers, which is usually achieved by applying the principle of quantum error correction~\cite{shor1995scheme,steane1996error}. For each known quantum error correction scheme, some quantum gates in the fault-tolerant universal set are easy to implement, while others are difficult. The latter can be efficiently executed with quantum software, which can be prepared in advance and then consumed during the execution of the quantum gate. The advantage of using software rather than hardware to implement the gate is that one can verify that the software is ready according to specifications before use. If a problem is found with the software, it can be discarded or repaired. On the contrary, if the hardware has a severe failure during the execution of the quantum gate, it may be difficult to recover. 

The idea for creating and preparing quantum states offline is not new. Instead, several researches~\cite{shor1994algorithms,knill1998resilient,gottesman1999demonstrating} have been carried out. Among them, Gottesman and Chuang~\cite{gottesman1999demonstrating} proposed an interesting approach to preparing, using, and classifying quantum states. The goal of their approach is to make the design of fault-tolerant quantum computers more straightforward and methodical. The main idea is to use a generalized form of quantum teleportation, a simple technique to reduce the required resources to construct necessary quantum gates (states), such as Pauli gates ($X$, $Y$ and $Z$), Cliff group, Toffoli gate, the $\pi$/8 gate (rotation about the $Z$-axis by an angle $\pi$/4), and the phase gate. Their construction relies on the ability to create some ancilla states, which is independent of the data being manipulated, and therefore can be prepared offline in advance. So they are valuable general-purpose quantum resources, perhaps a kind of "quantum software," as mentioned by Gottesman and Chuang~\cite{gottesman1999demonstrating}, and may be considered as commercial products that can be manufactured.

\begin{table*}[h]
\begin{threeparttable}
\centering
\caption{The Effects of QCS characteristics on Quality Attributes~\cite{sodhi2018quality}}
\label{table:QAs}
\renewcommand\arraystretch{1.1}
\small
\begin{tabular}{|l|p{0.3cm}|p{0.3cm}|p{0.3cm}|p{0.3cm}|p{0.3cm}|p{0.3cm}|p{0.3cm}|p{0.3cm}|p{0.3cm}|p{0.3cm}|} 
\hline 
{\bf QCS Characteristics}\hspace*{2.6cm}\rotatebox{90}{\tabincell{l}{\bf{Quality}\\\bf{Attributes}}} & \rotatebox{90}{Availability} & \rotatebox{90}{Interoperability} & \rotatebox{90}{Maintainability} & \rotatebox{90}{Manageability} & \rotatebox{90}{Performance} & \rotatebox{90}{Reliability} & \rotatebox{90}{Scalability} & \rotatebox{90}{Security} & \rotatebox{90}{Testability} & \rotatebox{90}{Usability}  \\
\hline  
{Platform heterogeneity} & U & $-$ & U & U & $-$ & U & $-$ & $-$ & U & $-$ \\
\hline
{Special physical environment} & U & $-$ & $-$ & U & $-$ & U & U & U & U & $-$  \\
\hline
{Large form factor} & $-$ & $-$ & $-$ & $-$ & $-$ & $-$ & U & $-$ & $-$ & $-$  \\
\hline
{Higher energy efficiency} & $-$ & $-$ & $-$ & $-$ & F & $-$ & F & $-$ & $-$ & $-$  \\  
\hline
{Lower level of the programming abstractions} & U & $-$ & U & $-$ & $-$ & U & $-$ & $-$ & U & $-$ \\
\hline
{Remote software development and deployment} & $-$ & $-$ & U & $-$ & $-$ & $-$ & $-$ & $-$ & U & $-$  \\
\hline 
{Dependency on quantum algorithms} & $-$ & U & U & $-$ & F & $-$ & F & $-$ & U & $-$  \\
\hline
{Limited portability of software} & U & U & U & $-$ & $-$ & $-$ & U & $-$ & $-$ & $-$ \\
\hline
{Limited quantum networking} & U & $-$ & $-$ & $-$ & U & U & U & $-$ & $-$ & $-$  \\
\hline 
{Lack of native quantum operating system} & $-$ & $-$ & $-$ & U & U & U & U & U & $-$ & $-$ \\
\hline 
{Fundamentally different programming model} & $-$ & U & U & $-$ & $-$ & U & $-$ & U & U & $-$ \\
\hline 
{Dependency on classical storage} & $-$ & $-$ & $-$ & U & U & U & U & $-$ & $-$ & $-$ \\
\hline
\end{tabular}
\begin{tablenotes}
    \item[$\ast$] Cell value indicates an impact on quality attributes: {\it "F":~Favorable, "U":~Unfavorable, "$-$":~Unknown/Neutral}.
\end{tablenotes}  
\end{threeparttable}
\end{table*}

\section{Quantum Software Measurement}
\label{sec:measurement}
Software metrics aim to measure the inherent complexity of software systems to predict the overall project cost and evaluate the quality and effectiveness of the design. Software metrics have many applications in software engineering tasks such as testing, maintenance, reengineering, reuse, and project management~\cite{pressman2010software,poulin1996measuring, fenton2014software,zuse2013framework}. 
Research for software measurement must adapt to the emergence of new software development methods, and metrics for new languages and design paradigms must be defined based on models relevant to these new paradigms ~\cite{bieman1996metric}. 
As research in quantum programming reaches maturity with a number of active research and practical products, software metrics researchers need to focus on this new paradigm to evaluate it in a rigorous and quantitative fashion~\cite{PiattiniPPHSHGP2020}. 

\subsection{Size and Structure Metrics}
\label{subsec:size-metric}

Zhao~\cite{zhao2021some} proposed some size and structure metrics for quantum software which mainly focus on measuring the size and structure of quantum software. Some of these metrics are the extensions of their classical counterparts, such as Lines-of-Code (LOC), Halstead’s Software Science, McCabe’s Complexity Metric, Henry and Kafura’s information flow Metric, while the others are specifically designed to quantify the quantum features in quantum software. These metrics are defined at different abstraction levels to represent various size and structure attributes in quantum software explicitly. The proposed metrics can be used to evaluate quantum software from various viewpoints. However, it is necessary to perform some experiments for evaluating the effectiveness of these metrics in practical quantum software development.

\subsection{Metrics for Quantum Circuits Understandability}
\label{subsubsec:circuit-metric}
Cruz-Lemus {\it et al.}~\cite{cruz2021towards}
proposed a set of metrics for accessing the understandability of quantum circuits. They defined these metrics from the various kinds of viewpoints, and grouped them by several categories: {\it circuit size}, {\it circuit density}, {\it single-qubit gates}, {\it multiple qubit gates}, {\it all gates in the circuit}, {\it oracles}, and {\it measurement gates}. The proposed metrics can be regarded as the first step to understanding the complexity of quantum circuits. 

\section{Empirical Study for Quantum Software Engineering}
\label{sec:empirical-study}
Although the emerging field of quantum software engineering (QSE) has recently attracted more attention, it is still unclear what the challenges and opportunities of quantum computing are facing the software engineering community. Recently, several researches have been carried out to address this issue from the perspective of empirical study. 

Shaydulin {\it et al.}~\cite{shaydulin2020making} surveyed open-source quantum computing projects from a different perspective, which mainly focused on the contributors of quantum software projects. They observed that one of the main problems in quantum computing is the lack of understanding of what training is required for success in the quantum computing field. To answer this question, they collected data on 148 contributors to three open-source quantum computing projects hosted on GitHub, including Qiskit~\cite{gadi_aleksandrowicz_2019_2562111} by IBM, PyQuil/Grove~\cite{smith2016practical} by Rigetti, and Cirq~\cite{cirq2018google} by Google. They studied the successful contributors to these projects from the field as well as the wider quantum computing community to understand the background and training that contributed to their success. These observations can help develop educational resources targeted at preparing a new generation of quantum computing researchers and practitioners. Their study could have a positive effect on bringing software engineering methodologies and techniques to the quantum domain. 

El aoun {\it et al.}~\cite{li2021understanding} conducted an empirical study on understanding the challenges faced by quantum program developers by analyzing QSE-related Stack Exchange forum posts and GitHub issue reports for quantum computing projects. To this end, they first qualitatively analyze the types of QSE-related questions on the Stack Exchange forum based on the existing taxonomy of question types on Stack Overflow and then use automatic topic modeling to reveal topics in QSE-related Stack Exchange posts and GitHub issue reports. Their study tried to answer three research questions:
\begin{itemize}
    \item What types of QSE questions are asked on technical forums?
    \item What QSE topics are raised in technical forums?
    \item What QSE topics are raised in the issue reports of quantum-computing projects?
\end{itemize}

The main findings of their study~\cite{li2021understanding} are
'\textit{QSE developers face traditional software engineering challenges (e.g., dependency management) as well as QSE-specific challenges (e.g., interpreting quantum program execution results). In particular, some QSE-related areas (e.g., bridging the knowledge gap between quantum and classical computing) are gaining the highest attention from developers while being very challenging to them.}’ 

Although the study is still primary, it may shed light on future opportunities in QSE, for example, to support the explanations of the theory behind quantum program code, and to interpret the results of quantum programs' execution.

\section{Software Engineering for Quantum Computing Platforms}
\label{sec:quantum-qcp}
A quantum computing platform is a complex software stack consisting of quantum computers or simulators on which a quantum algorithm could be executed~\cite{paltenghi2021bugs}. A quantum computing platform usually includes a quantum programming language, a compiler, and an execution environment that supports running quantum programs. Examples of quantum computing platforms include Qiskit by IBM, Cirq by Google, and Q\# by Microsoft. Recently, several pieces of research have focused on software engineering issues regarding quantum computing platforms. 

\subsection{\bf Quality Attributes on Quantum Computing Platforms}
\label{subsec:attribute-qcp}
With practical quantum computing systems (QCSs) becoming a reality rapidly, it is desirable to make use of their real potential in software applications. Therefore, it is crucial to determine the implications of QCSs for quantum software architecture, and some questions must be answered, like “{\it What are the key characteristics of QCSs, which are efficacious for quantum software development? }” and “{\it In what way does a QCS affect the quality attributes (and non-functional requirements) of quantum software applications? }”

To answer these questions, Sodhi~\cite{sodhi2018quality} presented an in-depth study on the state-of-the-art QCSs for identifying all such characteristics of a QCS that are relevant from a software architecture perspective, and performed an investigation on how these characteristics may affect the architecture of a quantum software application. As the first step, Sodhi sought and investigated related papers and software artifacts of QCSs and identified thirteen critical characteristics of QCSs. This includes platform heterogeneity, physical environment, large form factor, energy efficiency, lower level of the programming abstractions, remote software development and deployment, dependency on quantum algorithms, limited portability of software, limited quantum networking, lack of native quantum operating system, limited multitasking and multiprocessing, fundamentally different programming model, and dependency on classical storage. These key characteristics of QCSs form a base for further studies on how they could affect the quality attributes of the software architecture of a QCS. 

Quality attributes (QAs) are measurable or testable properties of a system that is used to indicate how well the system satisfies the needs of its stakeholders~\cite{bass2012software}. They, therefore, have a system-wide impact on the architecture, implementation as well as the operation of the system. 
For investigating the impact of QCS characteristics on the QAs, Sodhi~\cite{sodhi2018quality} adapted a slightly expanded list of QAs, which include availability, interoperability, maintainability, manageability, performance, reliability, scalability, security, testability, and usability. He only discussed the parts of a QA that are related to determine how this QA is affected by the characteristics of QCSs. To do so, one may take each characteristic of QSC and considers how it affects the various QAs. The impact of each characteristic on QAs is classified as {\it favorable}, {\it unfavorable}, or {\it neutral}. The summary of the impact of various characteristics of QCSs on QAs is shown in Table~\ref{table:QAs}, and the detailed discussions on those impacts can be referred to the paper~\cite{sodhi2018quality}. Additionally, several issues should be further studied, such as identifying more characteristics of QCSs to enhance the current list and investigating more QAs for obtaining more complete findings (insights) for the problem studied in the paper. In conclusion, Sodhi~\cite{sodhi2018quality} argued that the evolution of the technology would likely introduce additional concerns and factors that may affect the architecture of quantum software application, as quantum computing is undergoing rapid development.

\subsection{Bugs in Quantum Computing Platforms}
\label{subsec:bug-qcp}
With the continuous development of quantum computing, the importance of software platforms for developing quantum programs is also increasing. How to ensure the correctness of such a platform is becoming more and more critical. Therefore, it is needed to understand the bugs that usually appear thoroughly. Based on this, Paltenghi and Prade~\cite{paltenghi2021bugs} conducted an in-depth study on the bugs in the quantum computing platform. They selected 18 open-source quantum computing platforms as research objects, collected and checked a set of 223 real bugs. The research results show many of these bugs (39.9\%) are quantum-specific, and unique methods are needed to prevent and discover them. These bugs are distributed in different components, but quantum-specific bugs often occur in components that represent, compile, and optimize quantum programming abstractions. Many quantum-specific bugs manifest through unexpected output rather than apparent signs of wrong behavior, such as crashes. They also proposed a hierarchical structure of frequently wrong models, including ten new quantum-specific models. The findings by Paltenghi and Prade show the importance and universality of bugs in quantum computing platforms and can help developers avoid common mistakes. It also helps tool builders to deal with the challenge of preventing, discovering, and fixing these bugs.

\subsection{\textbf{Testing Quantum Software Stacks}}
\label{subsec:testing-qcp}
Wang {\it et al.}~\cite{wang2021qdiff} proposed QDiff, which is a novel differential testing approach for testing quantum software stacks (QSSes). The authors generated semantically equivalent programs and execute them. Then, they conducted statistical tests to compare results or report a crush to find errors. The authors evaluated QDiff with Qiskit, Cirq, and Pyquil. They used 6 seed quantum programs to generate 14799 program variants through the semantics-modifying and semantics-preserving source transformation process. Also, they used a filtering mechanism to speed up the execution. At last, QDiff found 6 instabilities beyond normal noise on these three widely used QSSes.

\section{Challenges and Opportunities}
\label{sec:challenge}

Quantum computing is an emerging field where new engineering methodology needs to be developed to address issues such as unforeseen changes in requirements, lack of expertise in software development and limited budgets that plague scientific software projects~\cite{shaydulin2020making}. The interdisciplinary nature of the quantum computing leads to the complexity of the field. This section discusses the challenges and opportunities in the area of quantum software engineering.

\subsection{Quantum Software Requirements Analysis}
\label{subsec:co-requirement}
Quantum software requirements analysis may include two stages: {\it stakeholder requirements definition} and {\it system requirements definition}. A key issue for the stakeholder requirements definition is to establish the critical performance requirements so that it is possible to determine whether quantum technology provides enough benefits over classical technology. Similarly, the acceptability (size, cost, capability) of the user system must be established as early as possible to determine the feasibility of using quantum technology in the application under consideration. Still, the types of benefits that quantum software systems may provide are not fully defined. Therefore, if the existing technology can meet the needs, especially considering that the technical risks of using quantum software systems may be high, users may not consider quantum software systems. However, research and expected efforts to utilize quantum systems in the coming years will likely provide the knowledge needed to define the needs of quantum systems properly so that they can be considered in the solution space. Therefore, in the process of defining stakeholder needs, it is necessary to ensure that quantum and non-quantum solutions can be adequately distinguished so as not to ignore quantum benefits.

The definition of system requirements is to transform the stakeholder requirements into a set of (technical) system requirements that the proposed solution must meet. The resulting system requirements cover functional, non-functional, performance, process, and interface aspects, and include design constraints. However, due to the limitations of the models, the definition of design constraints due to quantum technology effects might also be problematic. Therefore, one of the main problems in the system requirements definition process is whether the models used to establish the relationship and correlation between different technical requirements are adequate and sufficient.  This problem, however, can only be solved by developing better models and considering the effects of quantum technology adequately.

\subsection{Quantum Software Design}
\label{subsec:co-design}
 This section discusses challenges and opportunities regarding quantum software design from the perspectives of {\it architectural design} and {\it detailed design}.

\subsubsection{\textbf{Quantum Architectural Design}}
\label{subsubsec:co-architectural}
The software architecture of a system defines its high-level structure, revealing its overall organization as a set of interacting components~\cite{perry1992foundations,mary1996software}. A well-defined architecture allows an engineer to reason about system properties at a high level of abstraction. The incorporation of quantum computing effects on software architectures may require new methods to define the attributes, features, functions, and constraints assigned to architectural elements. This will require quantum software architects to reach a consensus and establish standard representations for quantum software components. The extent to which new types of components are introduced at the architecture level depends on the level of abstraction required, which is yet unclear. The functional architecture is unlikely to be affected by the introduction of quantum software components, though specific functions within the architecture may be new.

\vspace*{1.5mm}
\noindent
{\textit{$\bullet$ Quantum Architectural Patterns (Styles).}}\hspace*{1.3mm} 
The software architecture pattern provides a skeleton or template for the overall software architecture or high-level design of software applications~\cite{gomaa2011software}. Shaw and Garlan~\cite{mary1996software} mentioned the architectural style (or pattern) of software architecture, which is the repetitive architecture used in various software applications (see also~\cite{bass2000quality}). Architectural patterns can be classified into two main categories: {\it architectural structure patterns}, which focus on the static structure of the architecture, and {\it architectural communication patterns}, which focus on the dynamic communication among distributed components of the architecture. Currently, software architects do not have architectural patterns for quantum software systems, which can be used to adequately deal with the quantum issues when specifying the architecture of quantum software systems. The current repository for architectural patterns do not provide patterns which consider the quantum issues. Therefore, to perform the architectural-level design of quantum systems, it is necessary to identify some {\it quantum architectural patterns}, which are well-considered to model the quantum effect of the system.

\vspace*{1.5mm}
\noindent
{\textit{$\bullet$ Quantum Architectural Description Languages.}}\hspace*{1.3mm} 
Architectural description languages (ADLs)~\cite{clements1996survey} are formal languages that can be used to represent the architecture of a software system. They focus on the high-level structure of the overall application rather than the implementation details of any specific source module. Several ADLs have been proposed, such as Wright~\cite{robert1997formal}, Rapide~\cite{luckham1995specification}, UniCon~\cite{shaw1995abstractions}, Darwin~\cite{magee1996dynamic}, and ACME~\cite{garlan1997acme} to support formal representation and reasoning of classical software architectures. Recently, software architects do not have tools to adequately deal with the quantum issues when specifying the architecture of quantum software systems. Current ADLs do not provide primitives to specify quantum components and their connections with classical components. Using ADLs, software architects can specify the functionality of the system using components and the interaction between components using connectors. It is necessary to extend current ADLs to {\it quantum architectural description languages (qADLs)} to formally specify the architectures of quantum software systems for architectural-level design of quantum systems. We believe that such a qADL should contain at least those mechanisms, including specifications of classical components with interfaces and connections between interfaces (already provided in classical ADLs), specifications of quantum components, and specifications of connectors between classical and quantum components. Also, a qADL should support the tasks of formal analysis, verification, and validation of quantum software architectures.\\

\noindent
{\textit{$\bullet$ Software Quality Attributes.}}\hspace*{1.3mm} 
Software quality attributes~\cite{bass2000quality} refer to the non-functional requirements of software, which can have a profound effect on the quality of a software product. The software quality attributes of a system should be considered and evaluated during the development of the software architecture. These attributes relate to how the architecture addresses important non-functional requirements, such as performance, security, and maintainability. Other software quality attributes include modifiability, testability, traceability, scalability, reusability, and availability. 
As we summarized in section~\ref{sec:design}, Sohdi~\cite{sodhi2018quality} identified some critical characteristics of quantum computing systems, and studied how these characteristics may affect the quality attributes of quantum software architectures. A more significant issue worth investigating is how to use these quality attributes to evaluate the quality of the quantum software architecture during the architectural design.

\subsubsection{\textbf{Detailed Quantum Design}}
\label{subsubsec:co-detail}
The detailed quantum design provides information about the quantum software system as a whole and the individual quantum software components to enable implementation. An essential aspect of the detailed design is the selection of the technologies required for each quantum software component. Substantially, the inclusion of quantum software component increases the search space available to the quantum software engineer in making technology selection, but may lead to a challenge. The development of suitable models and their incorporation into quantum software design frameworks should be an area of intensive research effort in the future. Another important consideration is that at any particular stage in the quantum software life cycle, an appropriate level of quantum software modeling must be included. Moreover, it is crucial to recognize that the individual quantum software components may be developed by several or many different organizations. 

\vspace*{1.5mm}
\noindent
{\textit{$\bullet$ Quantum Design Patterns.}}\hspace*{1.3mm}
A design pattern describes a recurring design problem to be solved, a solution to the problem, and the context in which that solution works~\cite{gamma1995design,frank1996pattern}. Design patterns have been evolved over a long period and provide the best solutions to specific problems faced during software development. Learning these patterns helps inexperienced developers to learn software design in an easy and faster way. 
Design patterns have proved highly useful within the object-oriented field and have helped to achieve good design of applications through the reusability of validated components. These experiences tell us that it is crucial to identify the elements of good and reusable designs for quantum software and to start formalizing people’s experience with these designs through quantum design patterns. We hope that the quantum design patterns, if being identified, could benefit the quantum software development, and a quantum pattern catalog for quantum software could be especially useful. 

\subsubsection{\textbf{Design Models for Quantum Software}}
\label{subsubsec:co-model}
There is an absence of modeling techniques for quantum software system design. The current existing model of quantum software systems~\cite{Perez-Delgado2020quantum} is generally a simple extension of the classical modeling technique. The lack of suitable models is probably one of the most significant difficulties facing quantum software engineering, particularly as this may impact design, testing, and possibly maintenance parts of the quantum software life cycle. 

Design models for classical software engineering have been extensively studied from various perspectives. Among them, 
several notable design models are {\it data flow diagrams} (DFDs)~\cite{myers1978composite,yourdon1979structured}, {\it entity-relationship diagrams} (ERDs)~\cite{chen1976entity} and {\it unified modeling language} (UML)~\cite{boochunified}. We believe that the first natural step is to investigate these well-established design models to see if they could be extended to the quantum computing domain to support quantum software modeling by incorporating the quantum effects within the models appropriately. 

\subsection{Quantum Software Implementation}
\label{subsec:co-implementation}
Quantum software implementation refers to the development of quantum software entities that meet the requirements, architecture, and design. It is not uncommon to introduce new constraints in the implementation. These constraints must be reflected back into requirements. The essential design characteristics for achieving a subtle quantum effect are likely not to be compromised. Therefore, the process must include rules and tests to ensure that unwanted changes are not introduced. One can expect that reliability will become an important area of research in quantum software technologies. The principles derived from this research will provide a basis for the implementation strategy formulated during the implementation process. In short, the implementation process will need to incorporate the principles of quantum software reliability engineering. Testing the potential impact of implementation changes on the entire quantum software system requires modeling before agreeing to the changes.

On the other hand, it seems challenging to introduce new quantum programming languages into widespread practice. Perhaps a promising way is to define the requirements for the future high-level quantum programming, which may eventually lead to the development and widespread use of a more efficient quantum programming language. Another possible research direction could be to develop techniques and tools to generate quantum code from quantum software design specifications automatically.

\subsection{Quantum Software Reliability }
\label{subsec:co-testing}

This section discusses the challenges and opportunities on quantum software reliability, including fault model, testing, debugging, verification, and visualization. 

\subsubsection{\textbf{Fault Model for Quantum Software}} 
\label{subsubsec:co-fault}
In general, a fault model refers to an engineering model that may cause problems in the construction or operation of equipment, structures, or software~\cite{binder2000testing}. The unique features of quantum programming, such as superposition, entanglement, and no-cloning, do not occur in classical imperative, functional, or even multi-paradigm programming. Each feature can manifest new fault types that ultimately lead to quantum program failures. Therefore, to test quantum software, a fault model for quantum software is required. Such a fault model should be based on the peculiarities of quantum programs, and defined through careful analysis of the unique constructs of quantum programming languages and reflects an initial evaluation of classes of potential faults. Such a fault model can be used to measure the fault-detection effectiveness of automatic test generation and selection techniques of quantum software. Although work on identifying bug types for quantum software is just beginning~\cite{huang2018qdb,huang2019statistical}, more study should be carried out to build a practical fault model for supporting quantum software testing and debugging. 

\subsubsection{\textbf{Quantum Software Testing}}
\label{subsubsec:co-testing}
Systematic testing of quantum software systems must be based on fault models that reflect the structural and behavioral characteristics of quantum software. Criteria and strategies for testing quantum programs should be developed in terms of the fault model. As its classical counterpart, quantum software testing must explore the following issues rigorously and provide the answers for each of them to build effective and efficient testing frameworks for quantum software. 

\begin{itemize}
    \item How to define testing coverage criteria of quantum software?
    \item How to automatically and efficiently generate test cases for quantum software?
    \item How to evaluate the test data quality for quantum software?
    \item How to test quantum software regressively?
\end{itemize}   

Although some of these issues, such as testing covering criteria~\cite{wang2018quanfuzz} and test case generation~\cite{honarvar2020property}, have been addressed in current researches, it is far from practical uses for testing quantum software. 

\subsubsection{\textbf{Quantum Program Debugging}}
\label{subsubsec:co-debugging}
A commonly used classical debugging technique is the one in which a developer examines the program state through setting breakpoints~\cite{lazzerini1992program}. This technique, however, cannot be used to debug quantum software since any inspection of a quantum register can cause it to decohere~\cite{wolf2012artificial}. A simple variant of this technique that entails making a copies of the quantum register is similarly foiled due to the physical impossibility of making copies of quantum objects. One possible way is to use multiple quantum registers that are prepared in the same way. However, the probabilistic nature of quantum computation should serve as a constant reminder to the debugger of quantum software that no two quantum registers can be assumed to be identical. It is evident here that more investigations are needed.   
Recently, several pieces of research ~\cite{li2014debugging,huang2018qdb,huang2019statistical,liu2020quantum,zhou2019quantum,li2019poq} have been carried out that showed some promising initial results on debugging quantum software. However, it is still not clear what the appropriate debugging techniques for quantum computing are~\cite{mosca2019quantum}. As a result, new approaches still need to be developed. On the other hand, while assertion-based debugging~\cite{huang2018qdb,huang2019statistical,liu2020quantum,li2019poq} seems a promising way in debugging quantum software, different kinds of debugging techniques in classical computing, such as interactive debugging, static and dynamic analysis, code review and inspection, and post-mortem, are also worth further investigation~\cite{mosca2019quantum}. A question that naturally arises is: {\it are these classical debugging techniques applicable to the quantum domain?} 

As an example, {\it Whyline}~\cite{ko2008debugging}, a novel interactive debugging paradigm for classical software, can reduce the debugging time by allowing programmers to select “why” and “why not” questions to query the behavior of a program. We have not seen such a novel idea in the debugging of quantum software, but it would be worth exploring. An in-depth understanding of how quantum programmers perform debugging would provide an opportunity to develop new quantum debugging tools, and such work should be based on previous research that demonstrates best practices in scientific computing~\cite{ashktorab2019thinking}. 

Another example of classical debugging paradigms that should be paid attention to is {\it algorithmic debugging}~\cite{shapiro1982algorithmic,shapiro1983algorithmic}. Algorithmic debugging is an interactive process in which the debugging system gains knowledge of the expected behavior of the program being debugged and uses this knowledge to locate errors. An algorithmic debugger can be invoked after noticing an externally visible symptom of a bug. Then, it executes the program and builds an execution trace tree at the procedure level, while saving some useful trace information such as the procedure name and input/output parameter values. After that, the debugger traverses the execution tree and interacts with the user by asking for the expected behavior of each procedure. The user can answer “yes” or “no” to give an assertion about the predicted behavior of the procedure. Once some necessary conditions are satisfied, the search ends, and the location of the bug is identified. Algorithmic debugging provides a declarative way of debugging. If applied to quantum software, it might offer abstractions that allow programmers to think at an algorithmic level with less concern for details like control pulse generation, which may provide a possibility for overcoming the problems for which the quantum software debugging are facing~\cite{chong2017programming}.

\subsubsection{\textbf{Quantum Software Visualization}}
\label{subsubsec:co-visualization}
Understanding the behavior of quantum software is an important and nontrivial task, and software visualization~\cite{knight1999comprehension,stasko1998software}, a well-developed technique for understanding the behavior of classical software, may significantly contribute to it. Visualization of qubit states is particularly challenging since the number of achievable quantum states is exponential~\cite{ashktorab2019thinking}. Recently, several methods have been proposed for visualizing the state of qubits by using Bloch sphere~\cite{wikipedia2020bloch,gidney2017visualizing} and matrices of Pauli Expectation Value (PEVs)~\cite{wiseman1993interpretation,gross2010quantum}, and for visualizing the transitions between qubit states by using two-qubit representation~\cite{ibm2020entanglion}. However, the explosion of qubit state-space necessitates the development of scalable visualization that can intuitively help quantum software developers understand the states their systems are in, to facilitate the development, debugging, and validation of quantum algorithms~\cite{ashktorab2019thinking}.       

\subsubsection{\textbf{Quantum Program Verification}}
\label{subsubsec:co-verification}
Verification plays an essential role in quantum programming, both in the short and long term. Quantum programs are difficult to debug and test due to features such as superposition, entanglement, and no-cloning. To build the confidence regarding the correctness of a quantum program, we need to develop verification methods. There has been substantial progress in verifying quantum programs~\cite{chadha2006reasoning,brunet2004dynamic,feng2007proof,ying2012floyd,ying2013verification,rand2019formal,rand2018formally,liu2019formal}. However, as Rand {\it et al.}~\cite{rand2019formal} pointed out, novel verification methods are still needed to deal with errors and to verify error-prone quantum programs concerning the hardware we intend to run them on. Moreover, we need also approaches to verify quantum compilations~\cite{hietala2019verified}. 

\subsection{Quantum Software Maintenance}
\label{subsec:co-maintenance}
Software maintenance is an essential activity during software development. The objective is to modify the existing
software product while preserving its integrity~\cite{bennett2000software}. Methods, techniques, and tools for classical software maintenance have been well studied and established~\cite{bennett2000software,grubb2003software,dorfman1997software,sommerville2011software}, but research on the maintenance of quantum software systems is just starting~\cite{Castillo2020reengineering}. We believe that any quantum software maintenance process should deal with at least the following three main issues:

\begin{itemize}
    \item How to understand the existing quantum software?
    \item How to modify the existing quantum software?
    \item How to re-validate the modified quantum software?
\end{itemize}

Moreover, the maintenance process of quantum software systems likely needs to include the monitoring of the systems so that fault diagnosis could be informed during the maintenance and evolution of the systems.

\subsection{Quantum Software Reuse}
\label{subsec:co-reuse}
Component-based software engineering (CBSE) is an approach to software systems development based on reusing software components~\cite{kozaczynski1998component,heineman2001component}. Component here means a unit or a part of a model. It is the process that emphasizes the design and construction of computer-based systems using reusable software components or simply building software systems from pre-existing components~\cite{sommerville2011software}. The maintenance of CBSE can be easily accomplished by replacing or introducing a new component to the system~\cite{mili2001reuse}. 

As quantum software resources are getting accumulated, it is crucial to develop some methodologies, techniques, and tools for reusing quantum software systems. One promising way, we believe, might be the component-based quantum software engineering (CBQSE), which focuses on the design and development of quantum computer-based systems with the use of reusable quantum (classical) software components. Such an approach can save the development time and also be easy for maintenance during the evolution of the quantum software system. Another possible research direction, as we discussed in Section~\ref{subsec:co-design}, is to build a quantum architectural (design) pattern catalog to support the reuse of quantum software entities efficiently and effectively.

\section{Related Work}
\label{sec:work}

To the best of our knowledge, this work is the first comprehensive survey on the research of quantum software engineering regarding various phases of its life cycle, including quantum software requirements analysis, design, implementation, testing, and maintenance. This section discusses some related work in quantum programming languages, quantum software engineering, and quantum software development projects.

\subsection{Quantum Programming Language}
\label{subsec:QPL-survey}

Quantum programming languages play a vital role in the life cycle of quantum software development.
Several comprehensive surveys on quantum programming languages have been carried out from different perspectives.

Selinger~\cite{selinger2004brief} carried out the first survey on quantum programming language in 2004. His survey uses a similar classification scheme and offers a different perspective on some of the issues in the field. 

Gay~\cite{gay2006quantum} published a comprehensive and detailed survey on quantum programming languages in 2006. He classified the central theme of each paper from the perspectives of programming language design, semantics, and compilation. Regarding to programming language design, the paper considers imperative and functional programming languages, and for semantics, it focuses on the denotational techniques. Besides the survey, Gay also maintained an online bibliography as a resource for the community~\cite{gay2005bibliography}. 

Unruh~\cite{unruh2006quantum} also gave an overview of the state of the art on the development of quantum programming languages in 2006. In his overview, quantum programming languages are classified into two types, i.e., {\it practical programming language} and {\it formal programming language}. Practical programming languages, such as QCL~\cite{omer1998procedural,omer2000quantum} and Q~\cite{bettelli2003toward}, aim at practical applications such as simulation or the programming of actual quantum computers, while formal programming languages, such as QFC (QPL)~\cite{selinger2004towards,selinger2004brief,selinger2006lambda}, Quantum lambda calculus~\cite{maymin1996extending}, qGCL~\cite{sanders2000quantum,zuliani2001quantum,zuliani2001formal,zuliani2004non}, QPAlg~\cite{lalire2004process}, and CQP~\cite{gay2004communicating,gay2005communicating}, concentrate on how to model the semantics of a quantum program. The survey also pointed out some open problems and challenges regarding the practical and formal programming languages. For practical languages, the main challenges are how to design some powerful language constructs to abstract from the low-level model of quantum circuits to lift towards the high-level programming, and how to develop compilers and optimizers to obtain the best results out of the available resources. For formal languages, the main challenges lay in how to design expressive, easy-to-read quantum languages with well-defined semantics, and how to develop effective methods to verify quantum programs written by these languages.  

R\"{u}diger~\cite{rudiger2007quantum} presented a comprehensive survey for quantum programming languages from the pragmatic perspective in 2006. He first gave some introduction on necessary notations of quantum theory and quantum computation, and then discussed the goals and design issues for quantum programming languages. He claimed that quantum programming languages should enable programmers to reason about the structures of quantum algorithms and programs, and a well-designed quantum programming language should help find new quantum algorithms. The survey also discussed several concrete quantum programming languages such as pseudocode~\cite{knill1996conventions,knill2000encyclopedia}, QCL~\cite{omer1998procedural,omer2000quantum}, and Q language~\cite{bettelli2002architecture,bettelli2003toward} in detail, and gave a comparison of these three languages from the perspective of language paradigms and semantics. The survey also pointed out some research directions towards quantum programming language design and implementation.  

Jorrand's survey~\cite{jorrand2007programmer} on the quantum computing paradigm started from a brief overview of the principles that underlay quantum computing and introduced the breakthroughs achieved by the quantum computing field, which include quantum algorithms and teleportation. The main focus of the survey is on the quantum computation paradigm, that is, quantum programming languages and their semantics. Three styles for quantum programming languages have been discussed in the survey, including {\it imperative}, {\it functional}, and {\it parallel and distributed}. For each style, several languages are mentioned and compared, that is, QCL~\cite{omer1998procedural,omer2000quantum} and qGCL~\cite{sanders2000quantum,zuliani2001quantum,zuliani2001formal,zuliani2004non} for imperative, the approaches by Tonder~\cite{van2003quantum} and Girard~\cite{girard2004between} and QPL~\cite{selinger2004towards,selinger2004brief,selinger2006lambda} by Selinger for functional, and CQP~\cite{gay2004communicating,gay2005communicating} and QPAlg~\cite{lalire2004process} for parallel and distributed. The survey also discussed the semantics issues of quantum programming languages, from the perspectives of operational, axiomatic, and denotational aspects, respectively. 

Sofge~\cite{sofge2008survey} gave a brief survey on quantum programming languages from the perspective of history, method, and tools in 2008. He introduced a taxonomy of quantum programming languages, which divided into (1) imperative quantum programming languages, (2) functional quantum programming languages, and (3) other quantum programming language paradigms, which include some types of mathematical formalism that not intended for computer execution. Based on the taxonomy, a brief survey is given to reflect state of the art for quantum programming languages until 2007. Some challenges and a path forward for quantum programming languages are also given in the 
survey. 

Ying {\it et al.}~\cite{ying2012quantum} presented a survey on the programming methodology for quantum computing in 2012. Their survey discussed those issues regarding the design of sequential and concurrent quantum programming languages and their semantics and implementations, with more emphasis on the work conducted by the authors. Notably, in addition to discussion on the quantum programming languages themselves, they also reviewed some formal verification approaches for quantum programs and protocols, and discussed the potential applications of these quantum languages and verification approaches in quantum computing engineering.  

One of the most recent surveys for quantum programming languages is proposed by Garhwal {\it et al.}~\cite{garhwal2019quantum} in 2019, which gave an overview of state of the art in the field of quantum programming languages, focusing on actual high-level quantum programming languages, their features and comparisons. They classified quantum programming languages into five categories: quantum imperative programming language, quantum functional language, quantum circuit language, multi-paradigm language, and quantum object-oriented language. In the survey, they tried to answer the following research questions~\cite{garhwal2019quantum}: 

\begin{itemize}
    \item What are the different types of quantum languages? 
    \item What are the recent trends in developing quantum languages?
    \item What are the most popular publication venues for quantum programming?
    \item What are the most cited papers in the area of a quantum language?
    \item Which major companies, groups, institutes, and universities are working on the development of quantum languages? 
\end{itemize}

Another recent survey~\cite{zorzi2019quantum} is presented by Zorzi, which focuses on the QRAM (Quantum Random Access Machine) architectural model.
The survey is organized from the twofold perspectives, i.e., theory and concrete, and identifies the main problems one has to face in quantum language design. The survey tried to find some evidences from current researches to answer the following fundamental questions~\cite{zorzi2019quantum} regarding quantum language design and provided some possible answers to these questions.

\begin{itemize}
    \item What is the architectural model the language refers to?
    \item How to manage quantum data (which are no-duplicable to ensure the no-cloning property, so need a form of linear treatment)?
    \item What class of quantum functions one aim to program (expressive power)?
\end{itemize}

Besides those above, several other surveys are also given in~\cite{miszczak2011models,valiron2013quantum,valiron2015programming,hietala2016quantum,spivsiak2017quantum,chong2017programming}, on quantum programming languages from different viewpoints. However, although these surveys on quantum programming languages give very comprehensive pictures on the state of the art of researches on quantum programming languages, they lack the discussion of other phases of the life cycle of quantum software development, such as requirement, design, testing, and maintenance.  

\subsection{Quantum Software Engineering}
\label{subsec:QSE}
To the best of our knowledge, the term {\it quantum software engineering} was originally coined by John Clark and Susan Stepney~\cite{clark2002quantum} at the Workshop on Grand Challenges for Computing Research organized by Tony Hoare and Malcolm Atkinson, in 2002. Since then, there have been extensive research works on the different aspects of quantum software engineering as surveyed in this paper. Although no survey has been presented for quantum software engineering until now, some papers~\cite{stepney2005journeys,stepney2006journeys,barbosa2020software,PiattiniPPHSHGP2020,piattini2020quantum,gemeinhardt2021towards} discussed the challenges and opportunities of which quantum computing is facing on during quantum software development. 

In their seminal papers titled “Journeys in non-classical computation I and II”~\cite{stepney2005journeys,stepney2006journeys}, Stepney {\it et al.} presented a grand challenge on quantum software engineering, to develop a mature discipline of quantum software engineering for fully exploiting commercial quantum computer hardware. They claimed in ~\cite{stepney2006journeys} that “{\it The whole of classical software engineering needs to be reworked and extended into the quantum domain}.”, and introduced this challenge from different perspectives of quantum software engineering, including foundations, quantum computational models, languages and compilers, methods and tools, as well as novel quantum possibilities. Here, we briefly list some challenge problems for each aspect mentioned in ~\cite{stepney2005journeys,stepney2006journeys}:

\begin{itemize}
    \item {\it Foundations}: how to develop metaphors and models of quantum computing, which one can use to design and reason about quantum algorithms without considering the details within the quantum machine and unitary matrices, etc.  
    \item {\it Quantum computational models}: how to generalize various classical formalisms to the quantum realm in different ways
    \item {\it Languages and compilers}: how to determine the fundamental building blocks of quantum programming; how to design suitable assembly level and high-level quantum programming languages and the compilers of these languages; how to develop suitable reasoning systems and refinement calculi for these languages. 
    \item {\it Methods and tools}: how to develop powerful simulation systems so that one can perform computational experiments and validate the designs of the languages and algorithms; how to discover what high-level structuring techniques and architectures are suitable for quantum software; how to develop novel debugging and testing techniques for quantum computing, as we know that quantum execution is in principle unobservable.  
    \item {\it Novel quantum possibilities}: how to extend quantum software engineering to encompass those new issues from quantum mechanics, which cannot be even simulated by discrete deterministic classical computers.
\end{itemize}

We believe that the software engineering and the quantum computing communities should pay more attention to the issues listed above, to facilitate the construction of the overall picture for the field of quantum software engineering. 

Barbosa~\cite{barbosa2020software} presented a landscape of some research challenges to overcome for evolving the software engineering application in quantum computing, along with potential research directions in several aspects of the problem, including models, architectures, and properties. He believed that it is the time to discuss an agenda for a solid, rigorous software engineering discipline for quantum systems. He claimed that any roadmap for such a discipline should contain three main aspects: 

\begin{itemize}
\item How quantum software systems are modeled.
\item How the models of these systems are composed.
\item How the properties of these systems' behaviors can be predicted, specified, and verified.
\end{itemize}

He also pointed out some challenges and research directions regarding the models, architectures, and properties of the quantum systems, respectively. Among those directions, one particularly interesting issue, which is worth exploring, is how to extend the contract-based design, a successful paradigm in classical software engineering, to the quantum domain. The position paper is presented in which the formal aspect of the quantum software engineering principle is emphasized.

Piattini {\it et al.}~\cite{PiattiniPPHSHGP2020} presented the Talavera Manifesto for quantum software engineering and programming. The manifesto collects some principles and commitments about the field of quantum software engineering and programming, as well as some calls for action. They believe that quantum software engineering should have a necessary contribution to the success of quantum computing, and it is the time to develop quantum software by applying or adapting the well-established principles and methodologies from classical software engineering field, to the development of quantum software, which may include processes, methods, techniques, and practices. Concretely, they listed some principles and commitments regarding quantum software engineering, such as quantum software process, reengineering, requirement, evolution, testing and debugging, reuse, security and privacy, and management. 
In~\cite{piattini2020quantum}, they further gave a more detailed discussion about this manifesto with the emphasis on that "\textit{quantum computing will be the main driver for a new software engineering golden age during the present decade of the 2020s}." 

In the position paper~\cite{gemeinhardt2021towards}, Gemeinhardt {\it et al.} discussed the issue of Model-Driven Quantum Software Engineering, which combines the Model-Driven Engineering (MDE) with quantum computing to support quantum software development. To this end, they presented a research roadmap including several issues shown below, which should be considered when applying the MDE, particularly DSLs, for quantum computing. 

\begin{itemize}
    \item Modeling quantum data structures and algorithms
    \item Abstraction and exploitation of quantum technologies
    \item Deployment and integration via cloud-based quantum execution
    \item Executable DSLs for quantum computing
    \item Model-driven evaluation of quantum technologies
    \item Quantum computing for solving MDE tasks
\end{itemize}

They also discussed the challenges and opportunities for each issue. 


\subsection{Quantum Software Reliability}
\label{subsec:q-reliability}

Garc{\'\i}a de la Barrera {\it et al.}~\cite{garcia2021quantum} presented a comprehensive overview of the current state of the art on quantum software testing. The research aims to provide a general but formal overview of the current state of the art in quantum software testing and identifies which classic testing techniques could be applied to quantum software. This study attempts to answer the following research questions~\cite{garcia2021quantum}:

\begin{itemize}
\item What defects, limitations, and opportunities have been identified regarding the testing and verification of quantum software?
\item Which classic testing and verification techniques have been or could be adapted to quantum software?
\end{itemize}

The study also concluded several interesting trends in current quantum software testing, including (1) tracking uncertainty by repeatedly executing the algorithm and performing statistic approach over the measurements and (2) adaptations of Hoare logic by using quantum Hoare logic to reason about the correctness of quantum software. 

\subsection{Quantum Software Development Environment}
\label{subsec:IDE}

Roetteler {\it et al.}~\cite{roetteler2017design} presented a survey on investigating recent progress in building a quantum software framework, which can compile quantum algorithms from the high-level descriptions to physical quantum gates, which can be implemented on fault-tolerant quantum computers. In the survey, they discussed why tools such as compilation and design automation are essential to meet the enormous challenges of building scalable quantum computers. They also introduced a library developed with LIQUi|$\rangle$~\cite{wecker2014liqui} programming language, including reversible circuits for arithmetic and new real quantum methods that rely on quantum computer architecture, which allows the probability execution of quantum gates. This library, in some cases, can reduce time and space overhead. Also, the survey highlights why these libraries are useful for implementing many quantum algorithms. 

LaRose~\cite{larose2019overview} presented an overview and comparison of gate-level quantum software platforms. The overview mainly focuses on four new software platforms in quantum computing, namely, Forest (pyQuil) from Rigetti~\cite{smith2016practical,regetti2017forest}, Qiskit from IBM~\cite{ibm2017qiskit}, ProjectQ from ETH Zurich~\cite{projectq2017projectq,steiger2018projectq}, and Quantum Developer Kit (Q\#) from Microsoft~\cite{svore2018q}. Forest, Qiskit, and ProjectQ allow the users to access the real quantum computing devices, while Quantum Developer Kit only allows access to the simulator. In the overview, each platform is discussed and summarized from six aspects: requirements and installation, documentation and tutorials, quantum programming language syntax, quantum assembly/instruction language, quantum hardware, and simulator capabilities, with the links to documents and tutorial sources of each package. The overview also compares each platform with additional aspects, including library support, quantum hardware, and quantum compilers, and lists some notable and useful features of each platform. The purpose of this overview is to provide users with essential information for each platform to help them select a suitable platform to start exploring quantum computing programming. In addition to the four platforms mentioned, the overview also briefly introduced several quantum programming environments, including Quipper~\cite{green2013introduction,green2013quipper}, Scaffold~\cite{abhari2012scaffold}, and QCL~\cite{omer1998procedural}.

Fingerhuth {\it et al.}~\cite{fingerhuth2018open} presented an exhaustive review of open-source software projects in quantum computing. The review covers the whole stages in the quantum toolchain from quantum hardware interfaces through quantum compiler to implementations of quantum algorithms and also the full spectrum of quantum computing paradigms, which include quantum annealing~\cite{kadowaki1998quantum,finnila1994quantum,shin2014quantum}, and discrete and continuous-variable gate-model quantum computing~\cite{ortiz2017continuous}. For each project, the evaluation covers those features, including documentation, licensing, the choice of programming language, compliance with software engineering specifications, and the culture of the project. Fingerhuth {\it et al.} also discussed some of the findings that though the diversity of the projects is fascinating, only a few projects attract external developers, and even many commercially supported frameworks have flaws in software engineering. Based on these findings, the authors highlighted some best practices that could foster a more active community around quantum computing software, welcome newcomers to the field, and also ensure high-quality, well-documented code.

All these surveys focus specifically on the quantum software development environments, one of the essential aspects of quantum software development, but not on the whole life cycle of quantum software development, as we presented in this paper.  

\section{Concluding Remarks}
\label{sec:conclusion}
In its short history, quantum software development has been driven by the emergence of quantum programming languages~\cite{omer1998procedural,gay2005bibliography,green2013quipper,ibm2017qiskit,svore2018q}. In other words, quantum software development has been chiefly synonymous with quantum programming. While such languages have helped popularise the subject, this is not a healthy position in the longer term. In particular, it is vital that a complete software engineering discipline emerges for quantum software development. This paper is aimed as a stepping stone in this direction. This paper has examined state of the art in engineering support for quantum software systems, looking at requirements, design, implementation, testing, and maintenance in the quantum software life cycle. The crucial issues of quantum software reuse and measurement have also been considered. The evidence presented suggests that rapid progress is being made in these areas. However, it would be wrong to claim maturity in the topics addressed by this paper; indeed, although many techniques have emerged, further experience is required to appreciate their relative strengths and, indeed, to consolidate proposals into a small number of critical approaches.

\vspace*{3mm}
\noindent
{\large \bf Acknowledgements}

\vspace{1mm}
\noindent
The author would like to thank many colleagues for their valuable comments on the earlier draft of this paper and also for their continuing encouragement.

\bibliographystyle{ACM-Reference-Format}

\bibliography{qse-bibliography.bib}

\end{document}